\documentclass[english,aps,a4paper,floatfix]{revtex4}
\usepackage[T1]{fontenc}
\usepackage{tikz}
\usepackage{color}
\usepackage{amsmath}
\usepackage{amssymb}
\usepackage[latin1]{inputenc}
\usepackage{graphicx}
\usepackage{bm}
\usepackage{epsfig}
\usepackage{xcolor}
\definecolor{mynicegreen}{RGB}{102,182,102}

\begin{document}

\title{Cell theories for the chiral crystal phase of hard equilateral triangles}

\author{Yuri Mart\'{\i}nez-Rat\'on}
\email{yuri@math.uc3m.es}
\affiliation{Grupo Interdisciplinar de Sistemas Complejos (GISC), Departamento
de Matem\'aticas, Escuela Polit\'ecnica Superior, Universidad Carlos III de Madrid,
Avenida de la Universidad 30, E-28911, Legan\'es, Madrid, Spain}

\author{Enrique Velasco}
\email{enrique.velasco@uam.es}
\affiliation{Departamento de F\'{\i}sica Te\'orica de la Materia Condensada,
Instituto de F\'{\i}sica de la Materia Condensada (IFIMAC) and Instituto de Ciencia de
Materiales Nicol\'as Cabrera,
Universidad Aut\'onoma de Madrid,
E-28049, Madrid, Spain}

\begin{abstract}
	We derive several versions of the cell theory for a crystal phase of hard equilateral triangles. 
	To that purpose we analytically calculated the free area of a frozen oriented or freely rotating 
	particle inside the cavity formed by its neighbours in a chiral configuration of their orientations. 
    From the most successful versions of the theory we predict an equation of state which, despite being derived from a crystal configuration of particles, describes very reasonably the equation of state of the 
    6-atic liquid-crystal phase at packing fractions not very close from the isotropic-6-atic bifurcation. Also, the same equation of state performs well when compared to that from MC simulations for the stable crystal phase. The agreement can even be improved by selecting adequate values for the angle of chirality.
	Despite the success of two of the versions of the theory for the pressure, we show that the free-energy 
	is an increasing function of the angle of chirality, implying that the most stable phase is the achiral phase. Furthermore, we show that possible clustering effects, such as the formation of perfect chiral hexagonal clusters, which in turn crystallize into an hexagonal lattice, cannot explain the presence of the chirality observed in simulations.
\end{abstract}

\maketitle 

\section{Introduction}

The last two decades have witnessed the growth of theoretical and experimental works on the equilibrium liquid-crystal phase behavior of two-dimensional (2D) hard anisotropic particles. This effort is mainly motivated by the design, using micro-lithographic techniques, of prism-like particles with different cross-sectional polygonal geometries such as squares or rectangles, rhombuses, triangles and kites \cite{Zhao1,Zhao2,Zhao3,Zhao4}. These particles, after being adsorbed into monololayers, form effectively 2D systems of Brownian particles which approximately interact through entropic forces, making these systems ideal experimental platforms
to explore the effect of particle shape on the liquid-crystal symmetries in stable phases at equilibrium, and constitute a test bed for the vast number of theoretical works on 2D liquid crystals in the last years
\cite{Schlaken,Frenkel1,Dijkstra1,Frenkel2,MR1,Donev,MR2,Escobedo,Varga1,Varga2,Odriozola,Heras,Wittmann}.

Of particular interest are the symmetries of the liquid-crystal phases that can stabilized due to the specific particle shapes. For example, hard rectangles (and squares) of relatively small aspect ratio stabilize, before crystallization, a 4-atic (tetratic) phase with two equivalent, perpendicular directors. The orientational distribution function of this phase has a fourfold symmetry, $h(\gamma)=h(\gamma+\pi/2)$ \cite{Schlaken,Frenkel2,MR1,MR2}. For larger aspect ratios, rectangles undergo the standard isotropic 
(I)-nematic (N) transition before crystallization. Equilateral triangles stabilize a 6-atic 
phase with sixfold symmetry, $h(\gamma)=h(\gamma+\pi/3)$ \cite{Dijkstra2,MR3}. Hard-right triangles exhibit 
stable 4-atic liquid-crystal and crystal phases, but when the fluid is compressed, 
strong 8-atic orientational correlations are observed \cite{Dijkstra2,MR4}. We recently 
proposed a version of density-functional theory that captures the formation of clusters obtained from 
different monomeral triangular units, that in turn  explains the peculiar phase behavior of hard-right 
triangles \cite{MR5}.    

It has been shown recently that some 2D nonchiral particle shapes tend to generate chiral liquid-crystal or 
crystalline phases \cite{Dijkstra2,Zhao6,Zhao7,Zhao8,Carmichael}. For instance, triangular shapes with 
rounded corners have been shown experimentally to generate a chiral 6-atic liquid-crystalline phase \cite{Zhao6}. 
The presence of roundness and curvature in the edges and corners of the particles can promote certain configurations of dimers of particles that explain the presence of chirality \cite{Zhao6,Carmichael}. Note however that interactions other than entropic, such as electrostatic or polymer-mediated interactions,
could induce the observed chirality. Monte Carlo (MC) simulations on perfect hard equilateral triangles (HET) also stabilize chiral crystal phases: in equilibrium two coexisting domains, each one populated by one of two possible \emph{enantiomers}, contain equally oriented particles with respect to the main crystalline axes but with a nonzero chiral angle \cite{Dijkstra2}. The coexisting domains are separated by a non-chiral dividing interface \cite{Dijkstra2}. 
The chirality in the crystal phase was also found in NVT Monte Carlo simulations conducted by us in a fluid  of 
120-degree obtuse isosceles triangles \cite{MR5}. These triangles form equilateral triangular clusters, each composed of 
three monomeric units of original triangles. In turn the clusters crystallize into a triangular chiral lattice \cite{MR5}.

A physical explanation of chirality in the crystal phase of the purely entropic HET system is still lacking. A numerically implemented cell theory was recently used to explain chirality 
in the crystal phase of hard rhombuses \cite{Zhao8}, and it was shown that the entropic nature of interactions, 
together with the particular symmetry of the cell, can explain this phenomenon. 

In this work we develop different versions of cell theories for the crystalline phase of HET. We analytically calculate the free area of a given particle inside the cell formed by its neighbours in a particular chiral configurations. While the orientation of the neighbours is fixed, the \emph{caged} particle can freely rotate. We show that some of these versions predict an equation 
of state, not only for the crystalline, but also for the 6-atic liquid-crystal phase, in reasonable agreement with MC simulations. This fact shows that the liquid-crystal phase has a local spatial order similar to that of the crystal. By varying the degree of chirality (making it similar to that found in simulations) the agreement between theory and simulation can be improved. A very simple equation of state showing good agreement with MC results is obtained by assuming a frozen orientation for the caged particle, but with an extra global factor of $3/2$. We theoretically explain the origin of this factor.

By contrast, as far as the free energy is concerned, we show that the cell-theory free-energy at a fixed packing fraction is an increasing function of chirality angle, implying that the nonchiral phase is the most stable phase. This fact points to the presence of many-body correlations between particles, not included in the cell theory, that could explain the presence of chirality. The entropic self-assembly or clustering of 2D hard particles \cite{Glotzer1,Glotzer2,Glotzer3,Wittmann2} is a physical mechanism to explain the stability of some liquid-crystal symmetries, not obtained from the particle shape of the `monomeral' units of clusters \cite{MR5}.
Motivated by this result, we also explore the effect of clustering on the stability of the chiral phase by calculating the free area of hexagonal chiral cluster in the cell formed by its neighbours located 
in a hexagonal cell. We show that the free area is always lower than that of the perfect nonchiral hexagonal clusters, so that this particular arrangement of particles is not behind the observed chirality. 

The article is organized as follows. Section \ref{model} presents the model and the analytical expressions 
for the free energies and equations of state in their different versions, and make a comparison with MC simulations. Details on the analytical calculations are relegated to Sec. \ref{app}. In Sec. \ref{conclusion} some conclusions are drawn.

\section{Theoretical Model and Results}
\label{model}

\begin{figure}
        \epsfig{file=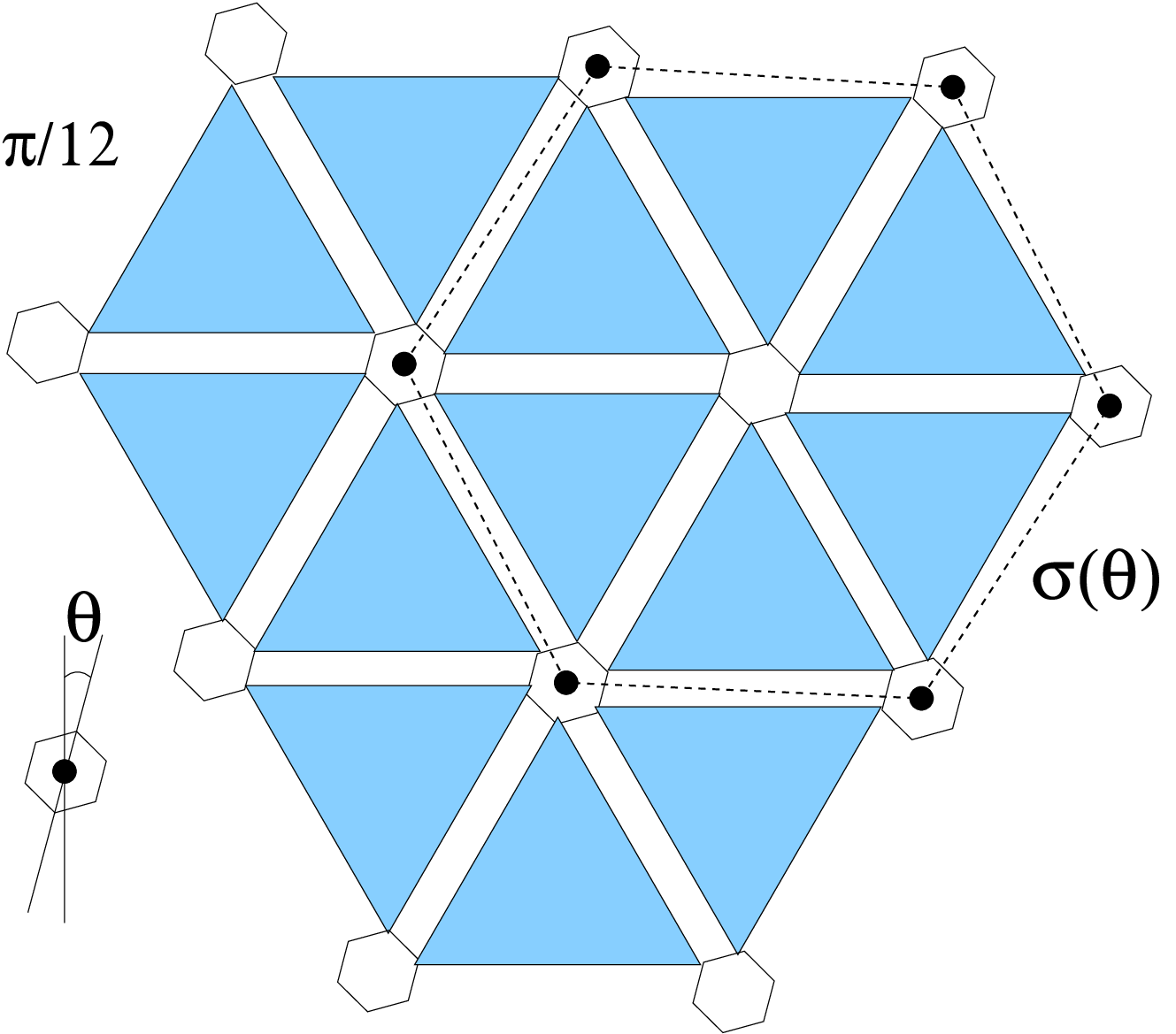,width=2.in}
        \epsfig{file=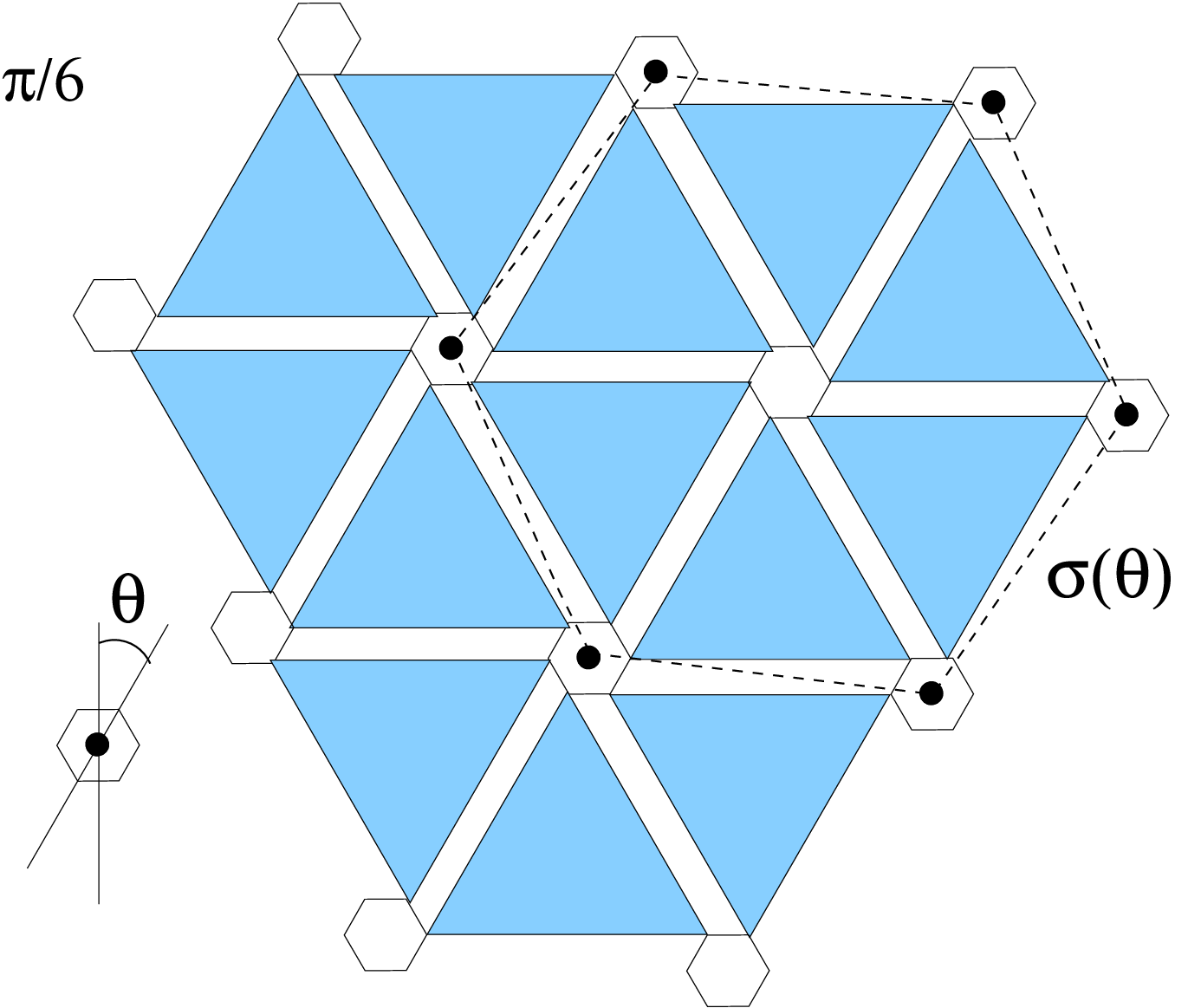,width=2.in}
        \epsfig{file=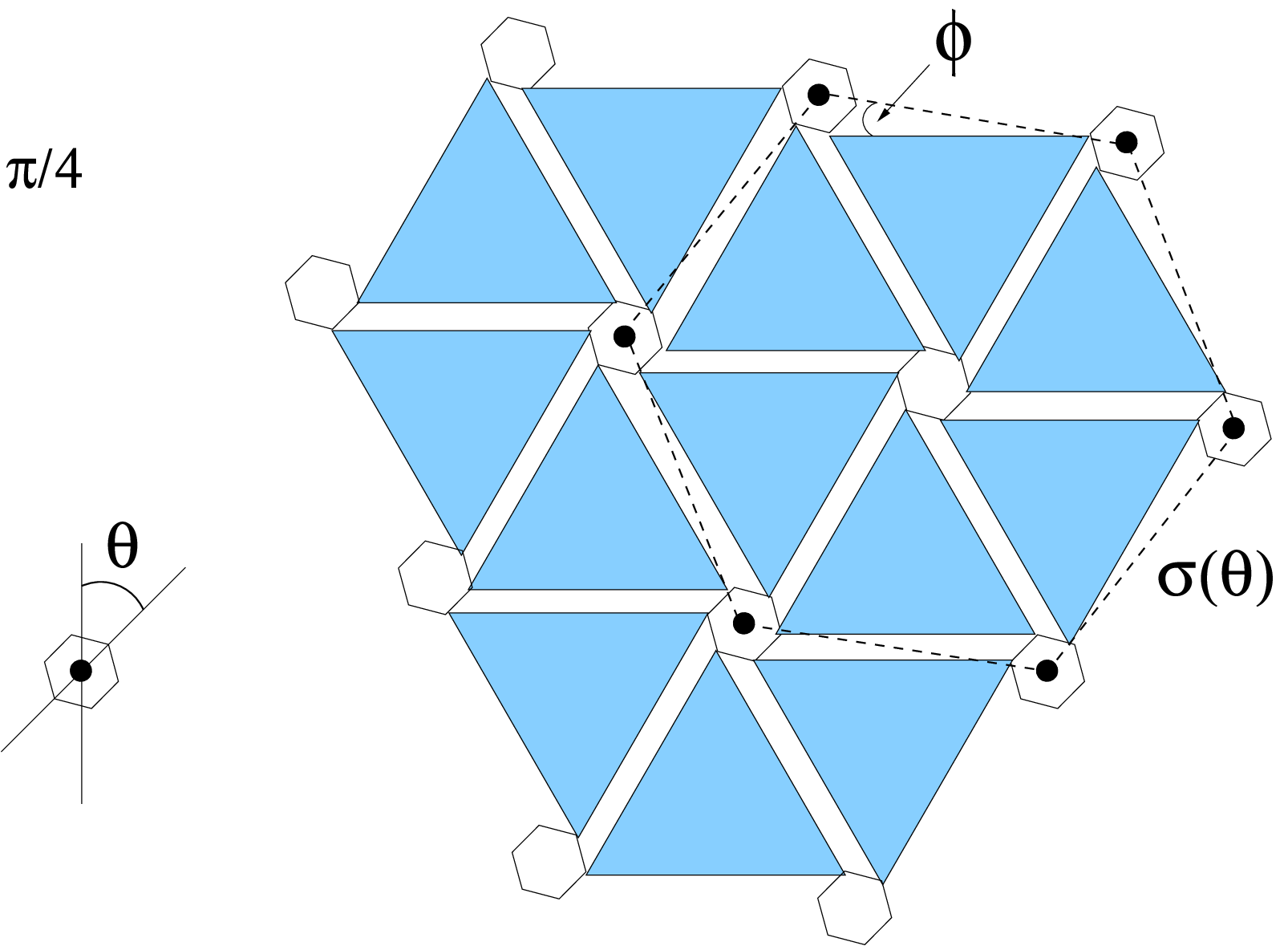,width=2.4in}
        \epsfig{file=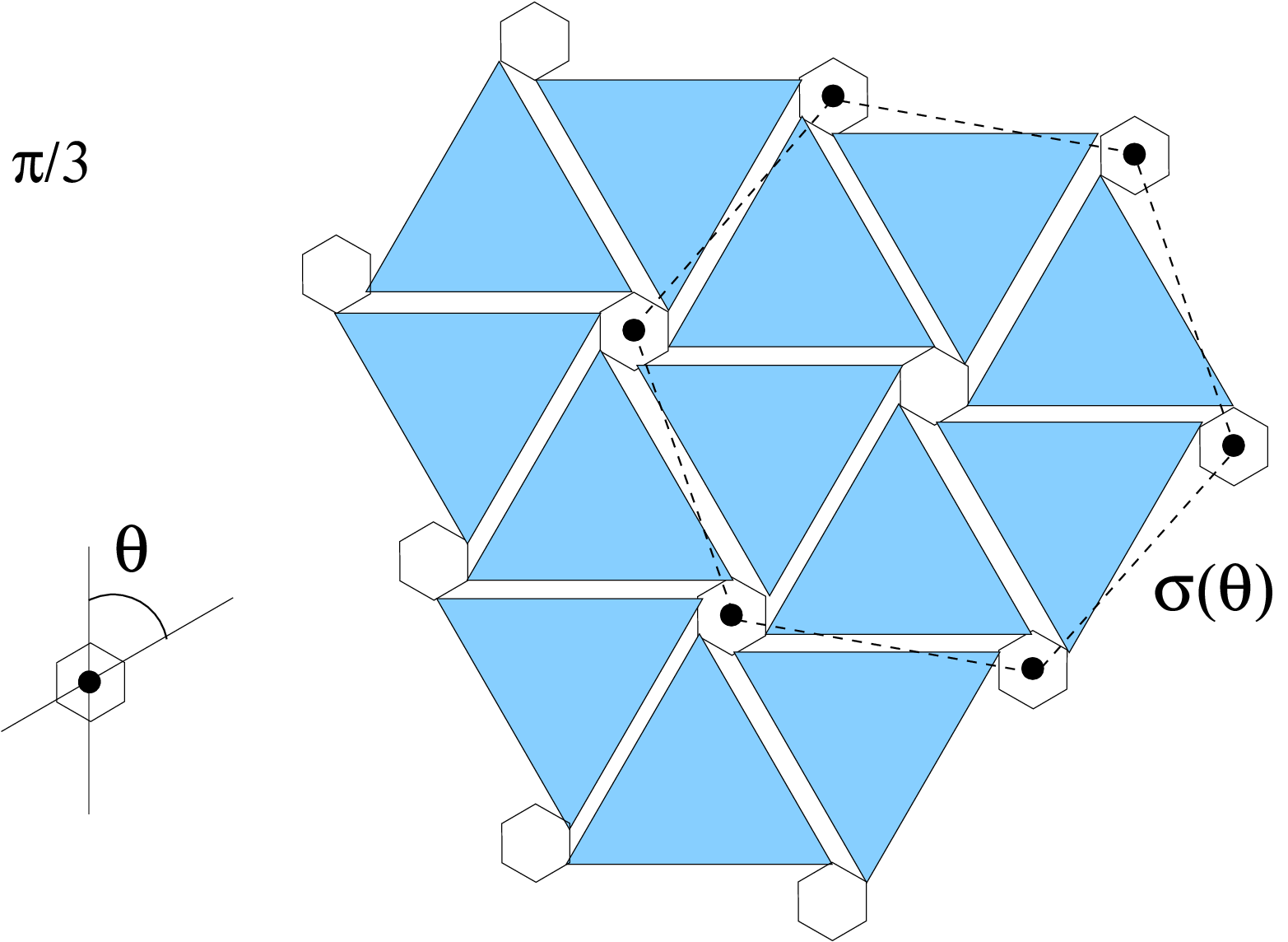,width=2.4in}
        \epsfig{file=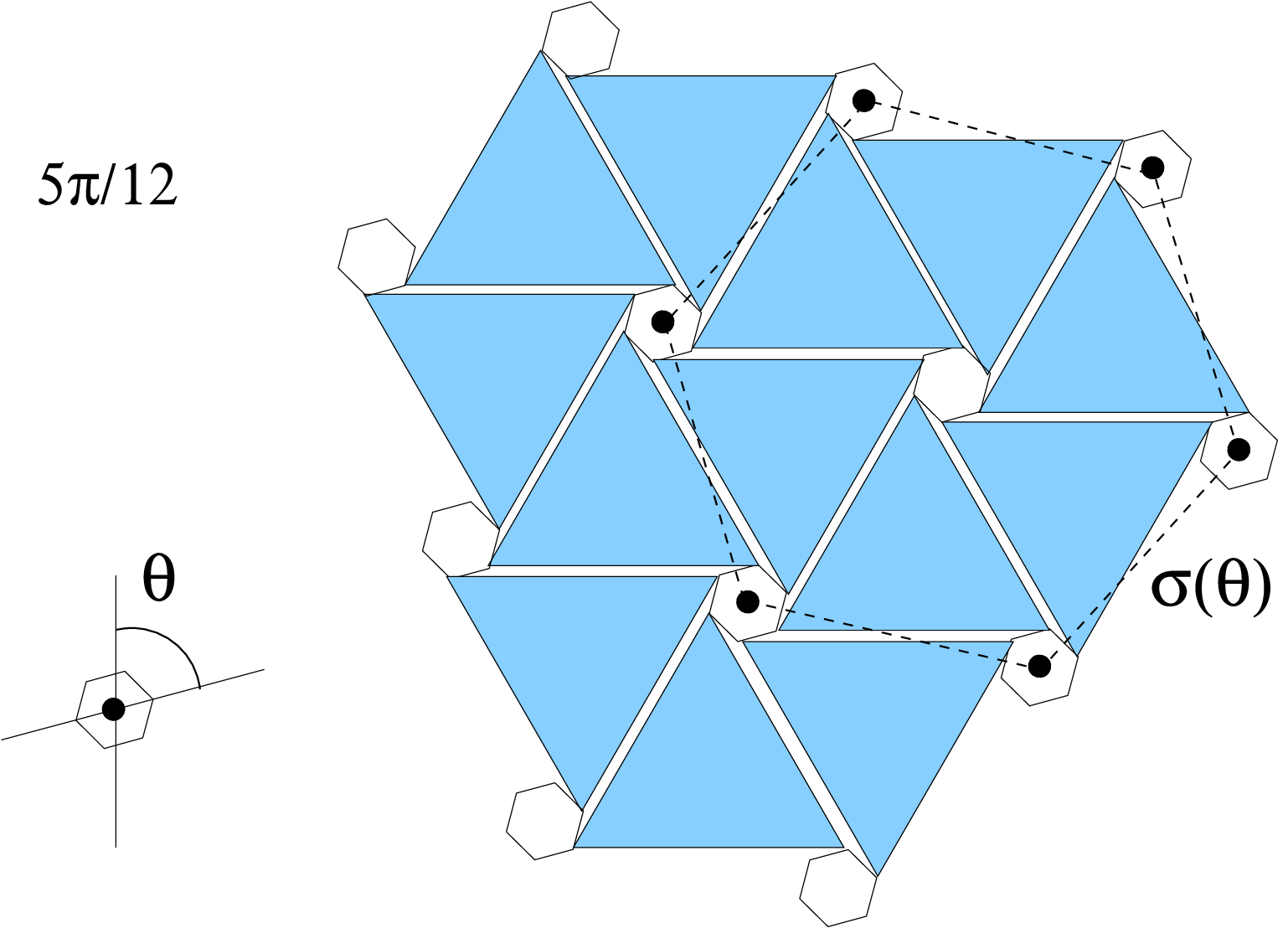,width=2.4in}
        \epsfig{file=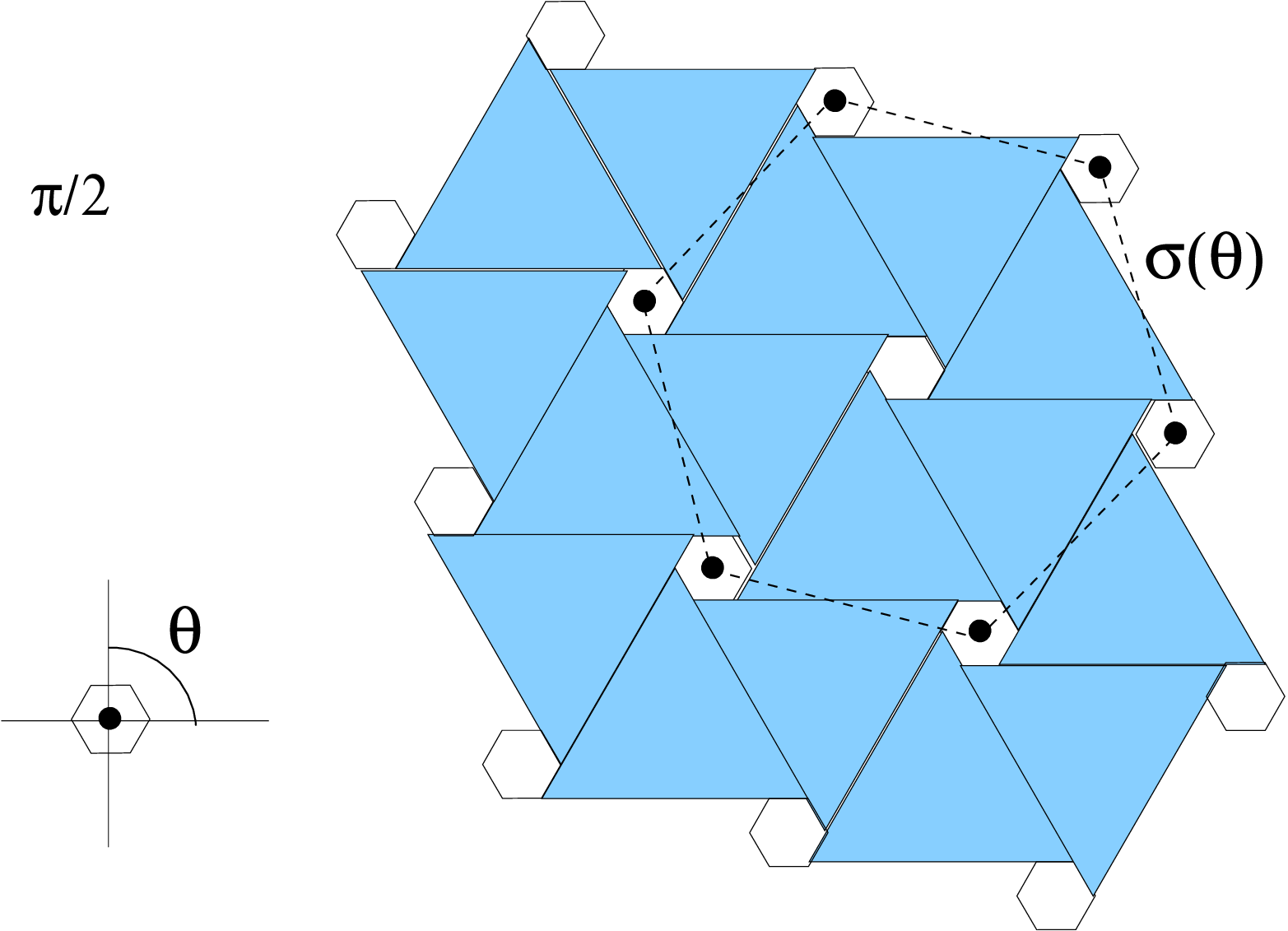,width=2.4in}
	\caption{Fragments of chiral lattices of HET with different values of chirality $\theta$ as indicated. 
	Small hexagons from which depart the triangular 
	vertexes are shown. The large hexagon (dashed line) of size $\sigma$ contain six triangles from which the packing fraction is obtained. 
	The angle $\phi$ between the triangular sides and that of the large hexagon is also indicated.}
	\label{fig0}
\end{figure}

The model for the crystal phase of HET is a simple 2D lattice where the centers of mass of triangles of side-length $l$ are located at the 
nodes of a triangular lattice. However we extend this model to include chirality, i.e. the rotation of the main particle axes with respect to the principal lattice directions. A chiral triangular lattice can be viewed as follows (a schematic is shown in Fig. \ref{fig0}). The vertexes of neighbour triangles are in contact 
with the vertexes of small hexagons. These hexagons are in turn rotated with respect to the triangle axes (which are fixed and parallel to each other) by an angle $\theta\in[0,\pi/2]$ (the chiral angle). 
This procedure is equivalent to keeping the hexagons fixed and rotating the triangles by the same angle. Note that the angle $\theta$ is not the one used to characterize chirality 
in MC simulations \cite{Dijkstra2}. Joining the centers of small hexagons we obtain a large hexagon (shown with dashed lines) inside which six triangles are located. The relative angle between the sides of triangles and those of the large hexagon (see Fig. \ref{fig0}) is called $\phi$; this is the angle used in Ref. \cite{Dijkstra2}.

\begin{figure}
	\epsfig{file=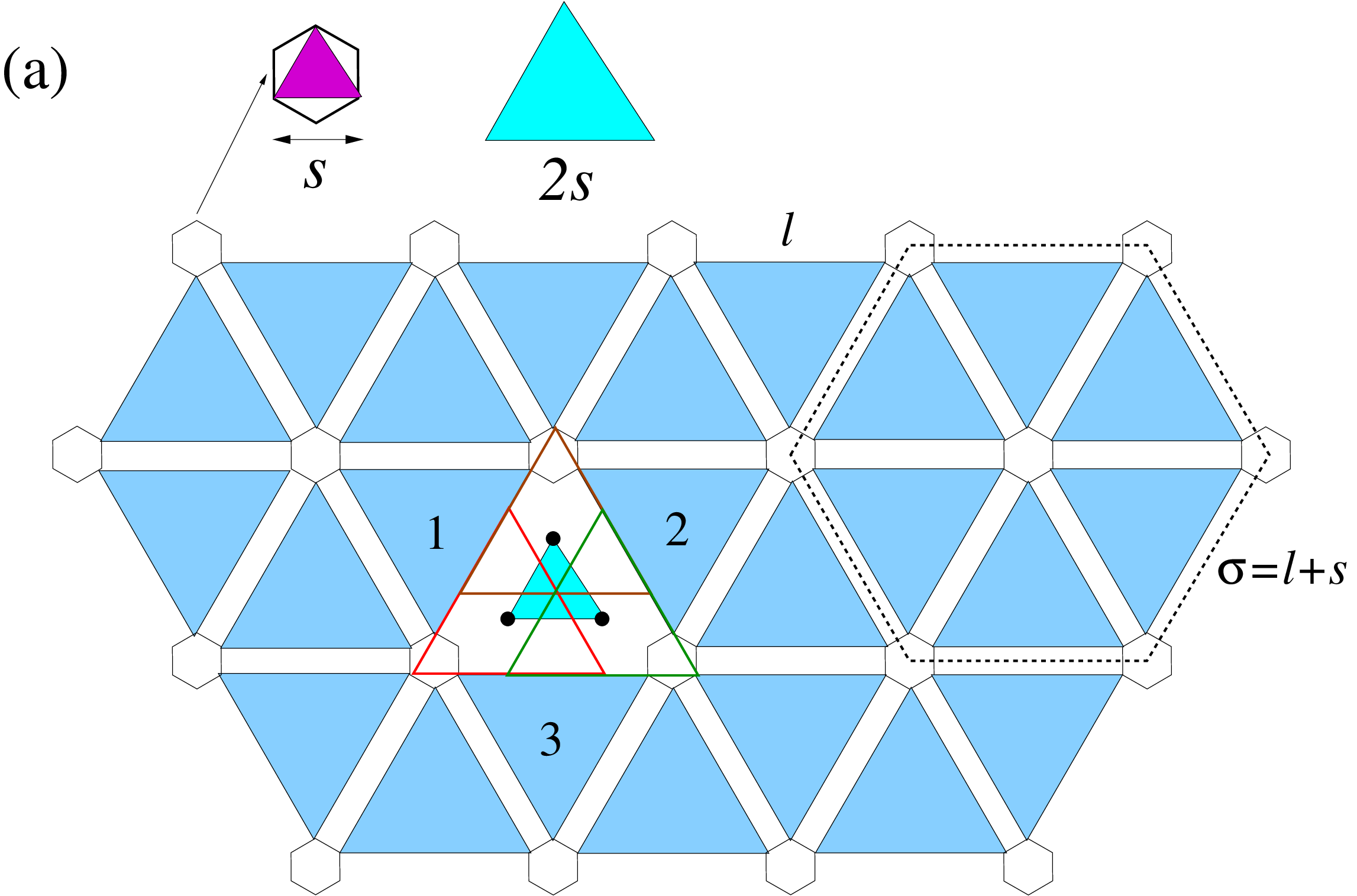,width=3.5in}
	\epsfig{file=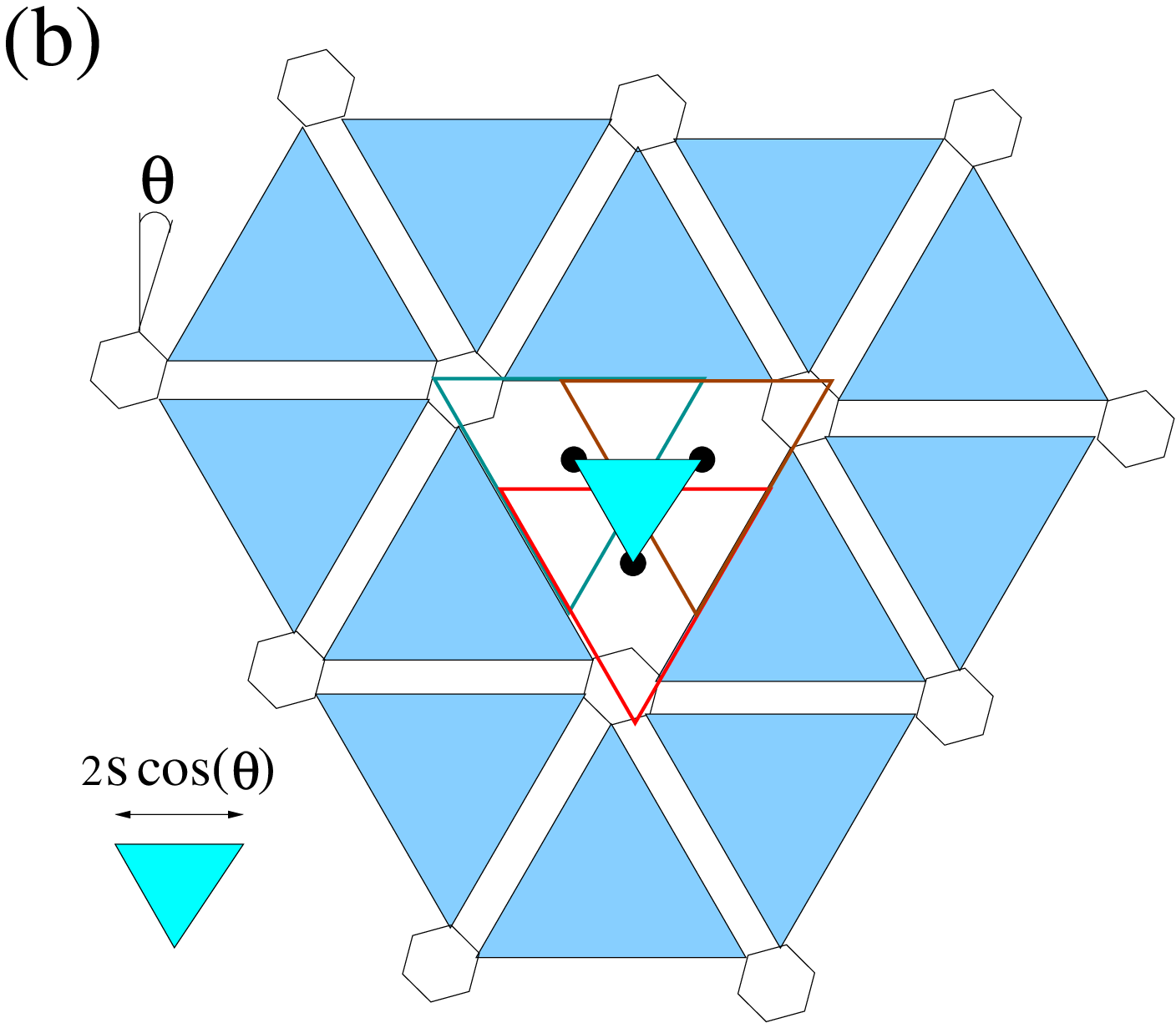,width=2.2in}
	\caption{(a) Fragment of the non-chiral ($\theta=0$) triangular lattice of HET where the size $s$ of the inscribed into small 
	hexagons equilateral triangle is 
	indicated. It is also shown the free-area (of size $2s$) of the caged triangle in the cell formed by its neighbours labeled as 1, 2 and 3. 
	(b): Fragment of the chiral ($\theta\neq 0$) triangular lattice of HET where the free area is also shown together with its size 
	$2s\cos\theta$.}
	\label{fig0b}
\end{figure}

First we formulate a very simple model: a cell theory with fixed orientation for all triangles. We calculate the free area of the center of mass of one triangle with the same (fixed) orientation as its neighbours 
inside the cavity formed by them. The neighbours are located at the nodes of the chiral lattice (as defined above). See Fig. \ref{fig0b} for the sketch of the free-area, which has the geometry of an equilateral triangle of side-length $2s\cos\theta$, where $s$ is the length of the (also equilateral) triangle
inscribed into the small hexagons (see Fig. \ref{fig0b}). Thus,
\begin{eqnarray}
	A_{\rm free}= \sqrt{3} s^2\cos^2\theta.
	\label{uno}
\end{eqnarray}
The side length $\sigma$ or the large hexagon as a function of $s$ and $\theta$ can also be calculated:
\begin{eqnarray}
	\sigma=\sqrt{l^2+s^2+2ls\cos\theta},
	\label{primera}
\end{eqnarray}
and from here the area of the large hexagon, $\displaystyle{a_{\rm hex}=\frac{3\sqrt{3}}{2}\sigma^2}$. The packing fraction is
\begin{eqnarray}
	\eta=\frac{6a_0}{a_{\rm hex}}=\left(\frac{l}{\sigma}\right)^2, \label{dos}
\end{eqnarray}
which is six times the area of one triangular particle, $\displaystyle{a_0=\frac{\sqrt{3}}{4}l^2}$, divided by the area of the large hexagon. 
Eqns. (\ref{primera}) and (\ref{dos}) allow to calculate $s$ in units of $l$ as a function of $\eta$ and $\theta$:
\begin{eqnarray}
	\tau\equiv \frac{s}{l}=\sqrt{\frac{1}{\eta}-\sin^2\theta}-\cos\theta, \label{lass}
\end{eqnarray}
and hence the free area $A_{\rm free}$ from Eqn. (\ref{uno}). 

\begin{figure}
	\epsfig{file=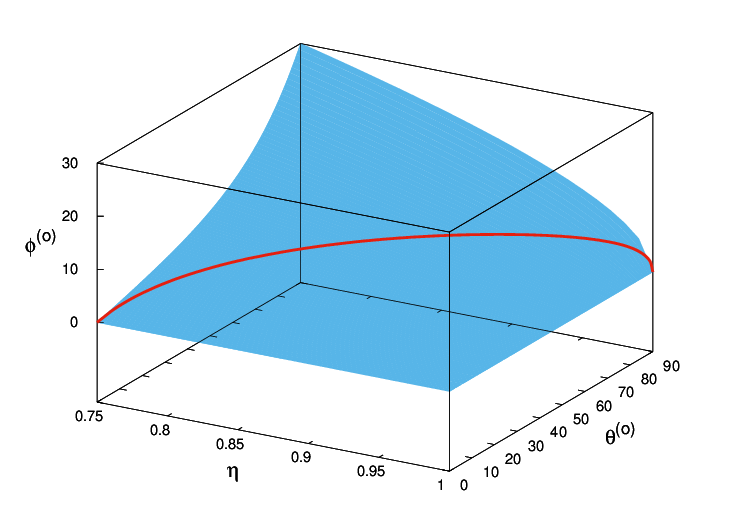,width=4.5in}
	\caption{$\phi$ as a function of $\theta$ and $\eta$. An example is shown of the path $\theta(\eta)=
	180\sqrt{\eta-0.75}$, where $\phi$ has a nonmonotonic behavior with $\eta$.}
	\label{nueva}
\end{figure}

The angle $\phi$ between the edge-length of the large hexagon and the triangular sides can be calculated as a function of $\theta$ and $\eta$ as
\begin{eqnarray}
	\cos\phi=\sqrt{\eta}\sin^2\theta+\cos\theta\sqrt{1-\eta\sin^2\theta},\quad 0\leq \phi\leq \phi_{\rm max}=\arccos\left(\sqrt{\eta}\right).
\end{eqnarray}
This angle will be used later to plot different quantities and to compare with MC results. Note that $\phi$ is a function of $\theta$ and $\eta$, and that $\phi\to 0$ when $\eta\to 1$ regardless of the value of $\theta$. This in turn means that even when chirality is large ($\theta\sim \pi/3$),
$\phi$ could be rather small if $\eta\sim 1$. Fig. \ref{nueva} is a 3D plot of the function $\phi(\theta,\eta)$. Note that, along particular curves $\theta(\eta)$, the angle $\phi$ could be a nonmonotonic function of $\eta$, exhibiting a maximum. This shows that the 
angle $\phi$ is not an appropriate measure of the strength of chirality, especially at high packing fractions.

Finally the cell-theory free energy per particle, using the parallel orientation approximation, is 
\begin{eqnarray}
	\varphi=-\ln\left(\frac{A_{\rm free}}{\Lambda^2}\right)=-2\ln\left(\frac{\sqrt[4]{3}l}{\Lambda}\right)
	-2\ln\left[\cos\theta\left(\sqrt{\frac{1}{\eta}-\sin^2\theta}-\cos\theta\right)\right],
	\label{simple}
\end{eqnarray}
with $\Lambda$ the thermal length. From the free energy, the pressure $p^*$ as a function of $\theta$ can be calculated. In reduced units,
\begin{eqnarray}
	p^*\equiv \beta p a_0=\eta^2\frac{\partial \varphi}{\partial \eta}=
	\left[\sqrt{\frac{1}{\eta}-\sin^2\theta}\left(\sqrt{\frac{1}{\eta}-\sin^2\theta}-\cos\theta\right)\right]^{-1}. \label{cuatro}
\end{eqnarray}
A much simpler expression results for $p^*$ as a function of $\phi$:
\begin{eqnarray}
p^*=\frac{\eta}{1-\sqrt{\eta} \cos\phi} \label{simple2}
\end{eqnarray}
As shown below, this equation of state performs poorly when compared with MC simulation results. Moreover the derivatives of $\varphi$ and $p^*$ with respect to $\theta$ are 
\begin{eqnarray}
	&&\frac{\partial\varphi}{\partial\theta}=2\sin\theta\left(\frac{1}{\cos \theta}-\frac{1}{\sqrt{\eta^{-1}-\sin^2\theta}}\right)\geq 0, \label{tres}\\
	&&\frac{\partial p^*}{\partial\theta}=-\frac{\sin\theta}{\left(\eta^{-1}-\sin^2\theta\right)^{3/2}}\leq 0.
\end{eqnarray}
Eqn. (\ref{tres}) in turn implies that chirality increases the free energy. Therefore, at this level of approximation, cell theory cannot explain the observed chirality in MC simulations.  

\begin{figure}
	\epsfig{file=fig4a.eps,width=3.in}
	\epsfig{file=fig4b.eps,width=3.in}
	\caption{Free area $A_{\rm free}/a_0$ (in units of the particle area $a_0$) as a function of $\theta$ (a) or $\phi$ (b) (measured in degrees) 
	from the freely rotating approximation of the caged triangle inside the cell formed by its neighbours in the chiral lattice (of chirality $\theta$ or $\phi$). 
	Results for three different values of packing fraction $\eta$ (as they are labeled) are shown. In the inset of panel (b) we show the 
	derivative of $A_{\rm free}/a_0$ with respect to $\phi$.
	}
	\label{fig1}
\end{figure}

The next step is to remove the constraint of a frozen orientation for the \emph{caged} triangle (that which wanders inside the cell formed by its neighbours).
The calculation of the free area is now cumbersome, and we relegate the details to the appendix \ref{app}. The procedure to calculate the free area is the following:
(i) Fix the angle $\gamma$ between the axis of the caged triangle and the rest of its neighbours, and move the center of mass of this particle, keeping contact with the neighbours. This step will define the envelope of the free area, ${\cal A}(\gamma)$, with a geometry in general more complicated than equilateral: For certain angular 
intervals of $\gamma$, the shape of the free area is made of an equilateral core plus three identical scalene triangles in contact with the equilateral core; while for other intervals the are reduces to a single  equilateral core. The areas of all of these triangles can be computed analytically as a function of $l$, $s$, $\theta$ and $\gamma$ (note that $s$ from Eqn. (\ref{lass}) is a function of $\theta$ and $\eta$). (ii) Analytically calculate the maximum allowed values of the angle $\gamma$, i.e. $\gamma_{\pm}$,  
following clockwise ($+$) and counterclockwise ($-$) rotations, for which the free area collapses to zero, 
${\cal A}(\gamma_{\pm})=0$. (iii) The  total free 
area can be computed via integration, $A_{\rm free}=3 \int_{-\gamma_-}^{\gamma_+} d\gamma {\cal A}(\gamma)$, which provides an analytical expression. The factor $3$ 
is related to the fact that, due to the symmetry of the equilateral triangle, the condition $\gamma\in[-\gamma_-,\gamma_+]$ reproduces only one third of the total free area. 
In Fig. \ref{fig1} the total free area $A_{\rm free}$ as a function of $\theta$ or $\phi$, and for different values of the packing fraction $\eta$, is plotted. Clearly $A_{\rm free}(\theta)$ is a decreasing function of $\theta$. Fig. \ref{fig1}(b) shows the derivative $\displaystyle{\frac{ d A_{\rm free}(\phi)}{d\phi}}$ vs. $\phi$. 
It can be seen that the derivative behaves linearly for angles $\phi\sim 0$, and consequently the function $A_{\rm free}(\phi)$ decreases quadratically for $\phi\sim 0$. 
The monotonically decreasing behavior of $A_{\rm free}$ means that the cell-theory free energy per particle, 
$\varphi=-\log \left(A_{\rm free}/\Lambda^2\right)$, is an increasing function of $\theta$ or $\phi$, which 
confirms that the new version of cell theory does not explain the stabilization of the chiral crystal. 

The total free area $A_{\rm free}$ (see Sec. \ref{app}) as a function of the small parameter $\tau$ can be Taylor expanded, which results in
\begin{eqnarray}
    \frac{A_{\rm free}}{a_0}=\frac{2}{\sqrt{3}}\left(2\tau\cos\theta\right)^3+o\left(\tau^3\right).
    \end{eqnarray}
Neglecting all terms beyond order $\tau^3$, and calculating the pressure using this approximation, we obtain the expression
\begin{eqnarray}
p^*=\frac{3}{2}\cdot \frac{\eta}{1-\sqrt{\eta}\cos\phi},
	\label{factor}
\end{eqnarray}
which is $3/2$ higher than the pressure obtained from the parallel particle approximation (see Eqn. (\ref{simple2}))  
and, as we will promptly see, it performs much better when compared with MC simulations.

In addition, an extension of cell theory to also include the average of the free area with respect to the angular distribution function $f_{\rm \theta}(\gamma)$ of the caged particle 
inside a chiral cell with chirality angle $\theta$ has been constructed. As usual, the total free energy has ideal and excess part contributions, 
\begin{eqnarray}
	\varphi\left[f_{\theta}\right]=\int_{-\gamma_-}^{\gamma_+}d\gamma f_{\theta}(\gamma)\ln f_{\theta}(\gamma)-1-\ln \frac{3\langle {\cal A}\rangle}{a_0},
	\quad \langle {\cal A} \rangle\equiv \int_{-\gamma_-}^{\gamma_+} d\gamma f_{\theta}(\gamma) {\cal A}(\gamma),
\end{eqnarray}
where we drop the trivial term $-\ln\left(a_0/\Lambda^2\right)$ and $\gamma_{\pm}>0$ denote the maximum 
rotation angles in the clockwise ($+$) and counterclockwise ($-$) directions. 
Functional minimization of $\varphi\left[f_{\theta}\right]$ with respect to $f_{\theta}(\gamma)$, 
with the constraint $\int_{-\gamma_-}^{\gamma_+}d\gamma f_{\theta}(\gamma)=1$, gives 
\begin{eqnarray}
	f_{\theta}(\gamma)=\frac{e^{{\cal A}(\gamma)/\langle {\cal A}\rangle}}{\int_{-\gamma_-}^{\gamma_+} d\gamma' 
	e^{{\cal A}(\gamma')/\langle {\cal A}\rangle}},
	\label{equil}
\end{eqnarray}
while the value of $\langle {\cal A}\rangle$ can be calculated self-consistently from its definition: 
\begin{eqnarray}
	\langle {\cal A}\rangle =\frac{\int_{-\gamma_-}^{\gamma_+} d\gamma {\cal A}(\gamma) e^{{\cal A}(\gamma)/\langle 
	{\cal A}\rangle}}
	{\int_{-\gamma_-}^{\gamma^+} d\gamma' e^{{\cal A}(\gamma')/\langle {\cal A}\rangle}}.
\end{eqnarray}
At equilibrium we find the free energy 
\begin{eqnarray}
	\varphi\left[f_{\theta}^{(\rm eq)}\right]=-\ln\left(\int_{-\gamma_-}^{\gamma_+} d\gamma e^{{\cal A}(\gamma)/\langle {\cal A}\rangle}\right) 
	-\ln\frac{3\langle {\cal A}\rangle}{a_0}, \label{energy}
\end{eqnarray}
and the pressure in reduced units
\begin{eqnarray}
	p^*=-\frac{\eta^2}{\langle {\cal A}\rangle}\left \langle \frac{\partial {\cal A}}{\partial\eta}\right\rangle.
\end{eqnarray}

\begin{figure}
	\epsfig{file=fig5a.eps,width=3.in}
	\epsfig{file=fig5b.eps,width=3.in}
	\epsfig{file=fig5c.eps,width=3.in}
	\epsfig{file=fig5d.eps,width=3.in}
	\caption{(a) Free area, averaged with respect to $f_{\theta}(\gamma)$, 
	as a function of (a) $\theta$ and (b) $\phi$ (both in degrees) for three different 
	values of packing fractions (as labeled). Free energy per particle $\varphi$ vs. (c) $\theta$ and (d) $\phi$ 
	at equilibrium for the same values of $\eta$. A constant term $-\ln\left(3a_0/\Lambda^2\right)$ was subtracted.} 
	\label{fig2}
\end{figure}

The averaged free area $\langle{\cal A}\rangle$ and free energy per particle $\varphi$, obtained from Eqn. (\ref{energy}), are plotted in Fig. \ref{fig2} (a) and (b) as a function of $\theta$ and $\phi$, respectively. 
The main difference between the values of $\langle {\cal A}\rangle$ and $A_{\rm free}$ (nonaveraged 
free area) which can be seen in Figs. \ref{fig1} and \ref{fig2},
is related to the fact that $A_{\rm free}$ is not weighted by the uniform distribution 
$\displaystyle{f_{\theta}(\gamma)=\frac{1}{\gamma_++\gamma_-}}$.
Although the free energy per particle, shown in Figs. \ref{fig2} (c) and (d),
is an increasing function of $\theta$ or $\phi$, and consequently chiral configurations 
are not entropically favored within the present theory,
we note that the slope $\displaystyle{\frac{d\varphi}{d\theta}}$ or 
$\displaystyle{\frac{d\varphi}{d\phi}}$ is very small, especially close to $\theta=0$ or $\phi=0$. 
This suggests that any hypothetical mechanism explaining chirality should involve small entropy gains overcoming 
the small entropy loss due to a smaller free area. 

\begin{figure}
	\epsfig{file=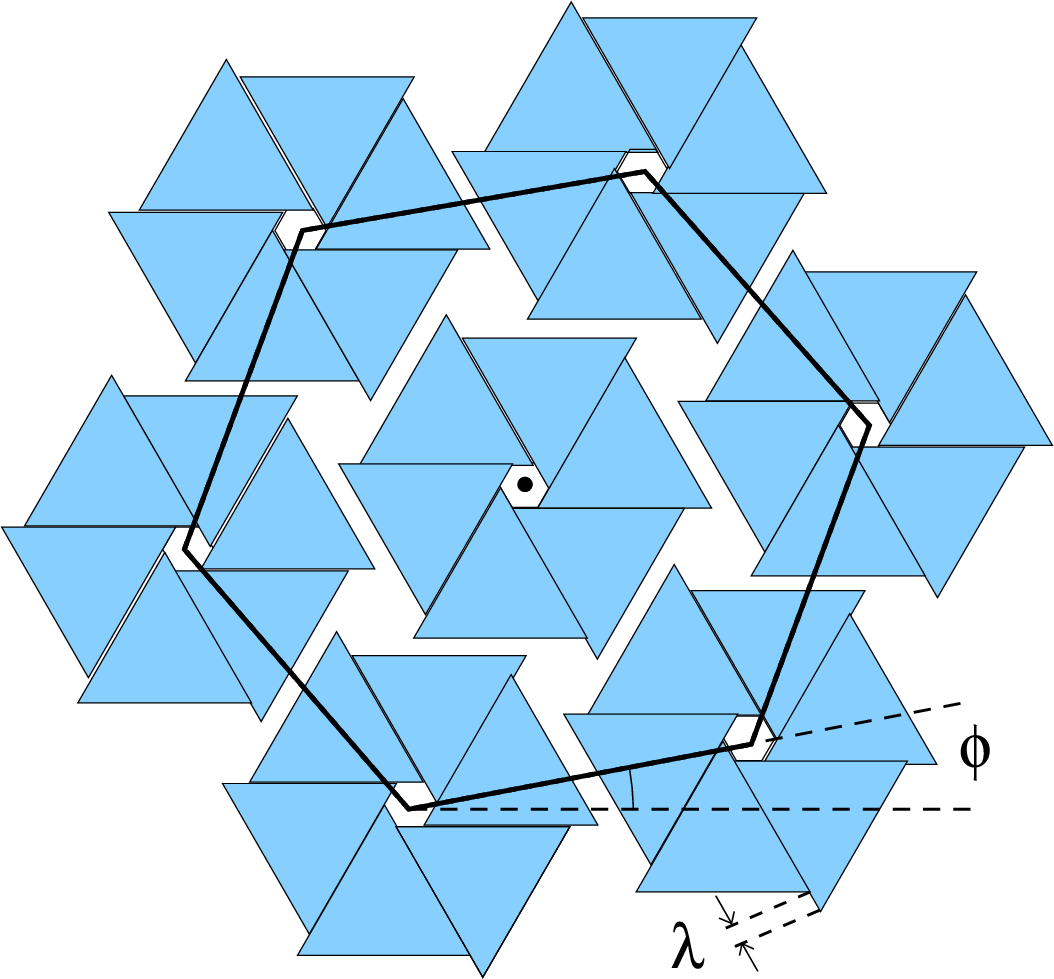,width=3.5in}
	\caption{Sketch of the hexagonal chiral clusters on a hexagonal lattice. The unit cell is drawn in
	thick solid line. The sliding length $\lambda$ and chirality angle $\phi$ are indicated.}
	\label{penultima}
\end{figure}

As an example, many body correlations 
could favor the presence of hexagonal clusters formed by six triangles in a chiral configuration, 
which crystallize into a hexagonal crystal phase (see Fig. \ref{penultima} for a sketch). We explored this avenue and conducted MC integrations to calculate the free area of chiral hexagonal clusters and their nonchiral versions (perfect hexagons), crystallizing into hexagonal lattices of the same packing fraction.  Fig. \ref{fig7_new} shows the values of the free areas for chiral
and nonchiral clusters made of six triangles, for different values of packing fraction and sliding length $\lambda$, 
and with $\phi=0$.
The parameters $\lambda$ and $\phi$ are indicated in Fig. \ref{penultima}. 
In particular $\lambda$  constitutes a free parameter to construct different configurations of the clusters for the same
chirality angle $\phi$ and packing fraction $\eta$. 
It can be seen that nonchiral structures made out of clusters always exhibit higher free areas than their 
chiral counterparts and the same behavior is obtained for $\phi\neq 0$. As a 
consequence, we can conclude 
at this point that particle clustering does not seem to be a viable mechanism to explain the chirality of the crystal phase of HET. We note that clustering structures other than hexagonal could be explored. However, MC simulations do not provide any hint as to alternative structures, as only pictures at very high density are provided in Ref. \cite{Dijkstra2}.
%The gain is the free-area of hexagons in 
%the hexagonal lattice 
%might compensate the lost in the free-area of a single triangle inside its hexagonal cluster. See a 
%further discussion about this point 
%in Sec. \ref{conclusion}.  

\begin{figure}
	\epsfig{file=fig7.eps,width=3.5in}
	\caption{Total free area (scaled with the 
	area $a=6a_0$), as obtained from MC integration, 
	of chiral and non-chiral clusters of six triangles as 
	a function of packing fraction for four different values of the sliding length $\lambda^*=\lambda/l$ 
	(as indicated in the box) and for $\phi=0$. Note that $\lambda^*=0$ corresponds to a perfect hexagonal cluster.}
	\label{fig7_new}
\end{figure}

\begin{figure}
	\epsfig{file=fig8a.eps,width=3.2in}
	\epsfig{file=fig8b.eps,width=3.2in}
	\epsfig{file=fig8c.eps,width=3.2in}
	\epsfig{file=fig8d.eps,width=3.2in}
	\caption{EOS of the chiral crystal phase of the fluid of HET from cell theories with chirality fixed to (a) $\theta=0^{\circ}$, (b) $30^{\circ}$,
	(c) $45^{\circ}$, and (d) and $60^{\circ}$. With symbols we show MC simulation results. The dotted line indicates the simplest cell theory model 
	with frozen orientation of the caged particle while solid and dashed lines (both very close to each other) are the results  
	from the cell theory of freely rotating caged triangle with uniform and minimized distribution functions $f_{\theta}(\gamma)$ respectively. The EOS from Eqn. (\ref{factor}) is shown 
	in dot-dashed line. For the 
 values of packing fraction in the range [0.70,0.89] MC simulations predict a stable 6-atic liquid-crystal phase with the 
 transition to the crystal phase at $\eta=0.89$ (labeled with arrow) \cite{Dijkstra2}. Thus our cell theory describes reasonably well also the EOS of the liquid-crystal phase for $\eta>0.75$ (not too close from the I--6-atic transition).
 }
	\label{fig3}
\end{figure}

Despite this failure, the two theories presented (namely freely rotating caged particle with uniform orientational distribution or else obtained via functional minimization) enjoy an important success: their fair performance 
%But the major success of both theories: freely rotating probe triangle with its distribution function $f_{\theta}(\gamma)$ being uniform or obtained via functional minimization 
%are their great performance 
in describing the equation of state (EOS) obtained from MC simulations. Fig. \ref{fig3} compares both theories, together with that derived from the frozen orientation approximation (see Eqn. (\ref{simple})). The free-rotation approximation accurately describes the EOS in an interval of packing fraction from 0.75 to 0.9, while 
the frozen-orientation approach provides a poor representation. 
In fact is is possible to optimize the value of $\theta$ (or $\phi$) for which the theoretical description is better. Fig. \ref{fig3} allows to discard values $\theta> 45^{\circ}$, while values $\theta\sim 30^{\circ}$ (or $\phi\sim 3.44^{\circ}$, 
$2.55^{\circ}$ and $1.68^{\circ}$ for 
$\eta=0.8$, 0.85 and 0.9 respectively) seem to optimize the agreement with simulations \cite{Dijkstra2}. Actually, as simulations show, in the liquid-crystal interval of packing fractions, the chirality of the fluid is completely absent. Chirality emerges beyond a particular packing fraction in the crystal phase. From Fig. \ref{fig3} we can infer that, from $0.75$ 
up to $0.87$, the best fitting from cell theory is obtained for $\theta=0$, while beyond $0.87$ a chirality angle between $30^{\circ}$ and $45^{\circ}$ improves the description of the EOS in the crystal phase. Unfortunately, the dependence of chirality on packing fraction cannot be obtained in the context of our theory, due to the monotonically increasing 
behavior of the free energy as a function of $\theta$.

\section{Discussion and conclusions}
\label{conclusion}

We have proposed three versions of cell theory for the chiral crystal phase of HET. 
The simplest theory involves the constraining of the caged particle inside a cell 
defined by the fixed positions of its neighbours, located at the nodes of the chiral triangular 
lattice, and with their axes parallel to each other. Moreover the orientation of the caged particle 
is identical to those of its neighbours. We have shown that this version fails to predict the EOS of the liquid crystal and crystal phases obtained from MC simulations. However, when multiplied by a factor of $3/2$, whose
origin comes from the leading order in a small lattice-parameter expansion of the free area for a freely-rotating caged particle, the performance in the description of the EOS greatly improves.

We derived a second version of the theory by removing the constraint of frozen orientation of the caged particle (keeping the orientation of its neighbours frozen) and allowing it to rotate inside the cell with a uniform 
angular distribution function. We have calculated analytically the integral of the free area over the angular degrees of freedom and the associated free energy. The ensuing pressure considerably improves the EOS  
when compared with MC results pertaining to the EOS of the 6-atic liquid-crystal and crystalline phases.   
By optimizing the chiral angle one can improve the agreement with MC simulations, the conclusion being that 
the chiral angle cannot be larger than $45^{\circ}$, also in agreement with simulations. 

The third and last version of the theory involves the averaging of the free area over the angular distribution function. This function is obtained from minimization of a proposed free-energy density functional consisting
of ideal angular term plus excess part, approximated as minus the logarithm of the averaged free area, 
which has been calculated numerically using the Newton-Raphson method. The EOS obtained from this version is practically indistinguishable from that obtained with the previous theory.

We have shown that the free energy per particle from the three theories are always increasing functions 
of the chiral angle, implying that the most stable configuration of the particles are achiral. However, 
a more detailed study of the integrated free area with respect to clockwise and counterclockwise rotations (assuming a clockwise chirality), shows that the former exhibits a maximum at a very small chiral angle. This fact, and the small value of the slope of the free-energy as a function of the chiral angle, makes it plausible the existence of another mechanism, based most likely on many-body particle correlations, with an associated increase in entropy, sufficient to overcome the entropy loss involved in a chiral rotation. Exploring this avenue will have to wait until a density-functional theory for inhomogeneous densities, incorporating exact correlations beyond two particles, can be developed.
Looking for a possible mechanism to explain chirality we proved that, in the context of cell theory, there is no free-area gain when particles are clustered into perfect chiral hexamers, as compared with their nonchiral counterparts: this particular clustering effect is not behind the chirality of the crystal phase observed in MC simulations. 
%Groups of six of triangles could adopt configurations 
%close to the chiral hexagonal clusters with their centers of mass positioned on the hexagonal lattice. The 
%gain in the free-area of these chiral hexagonal clusters could be higher than that of the achiral clusters (perfect 
%hexagons) resulting in the stability of the chiral configuration of particles. This conjecture can be checked via 
%MC simulations, something that we leave for a future work.

\begin{appendix}
	\section{Free area of the freely rotating caged triangle}
	\label{app}

	\begin{figure}
		\epsfig{file=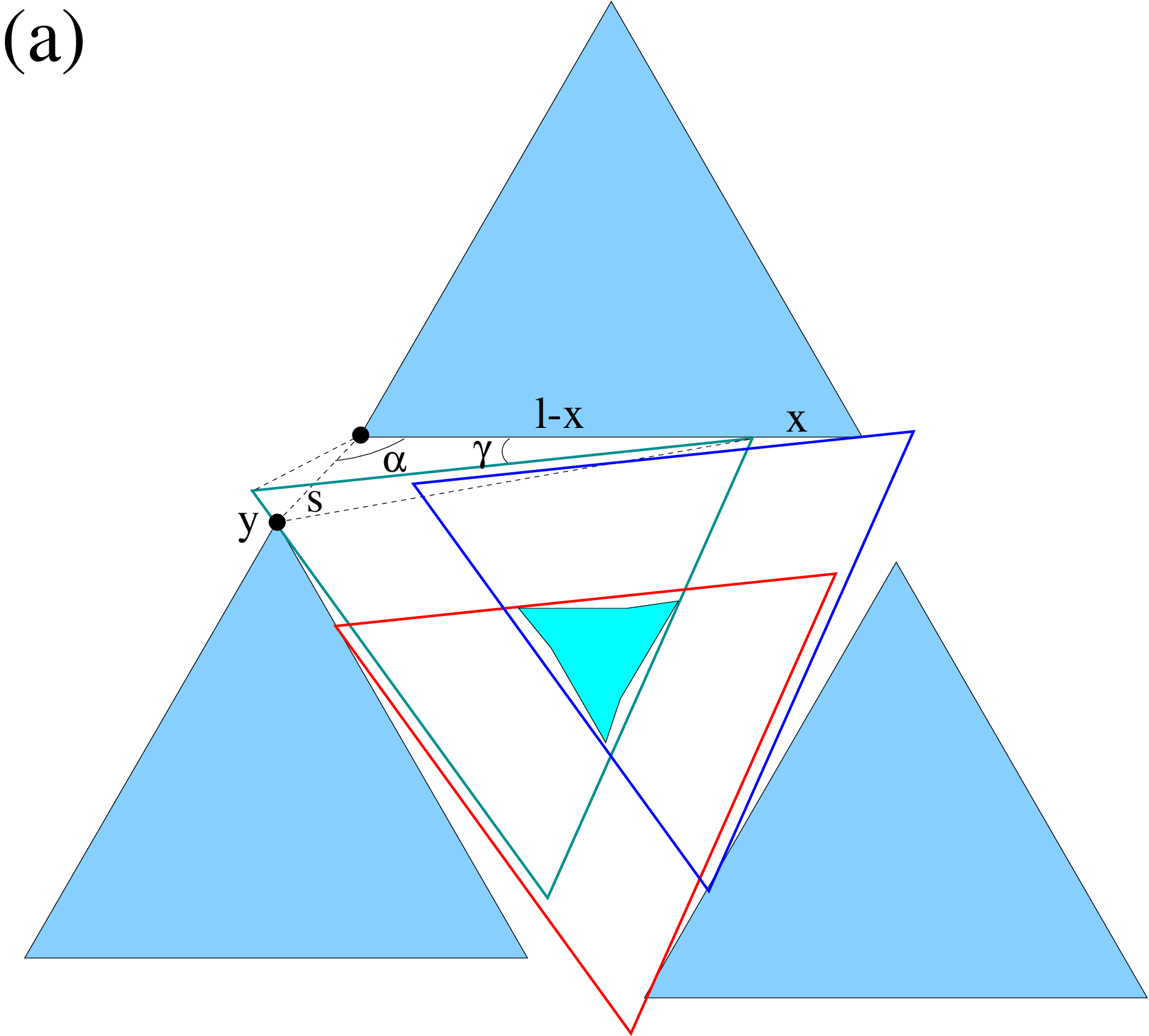,width=2.7in}\hspace*{0.5cm}
		\epsfig{file=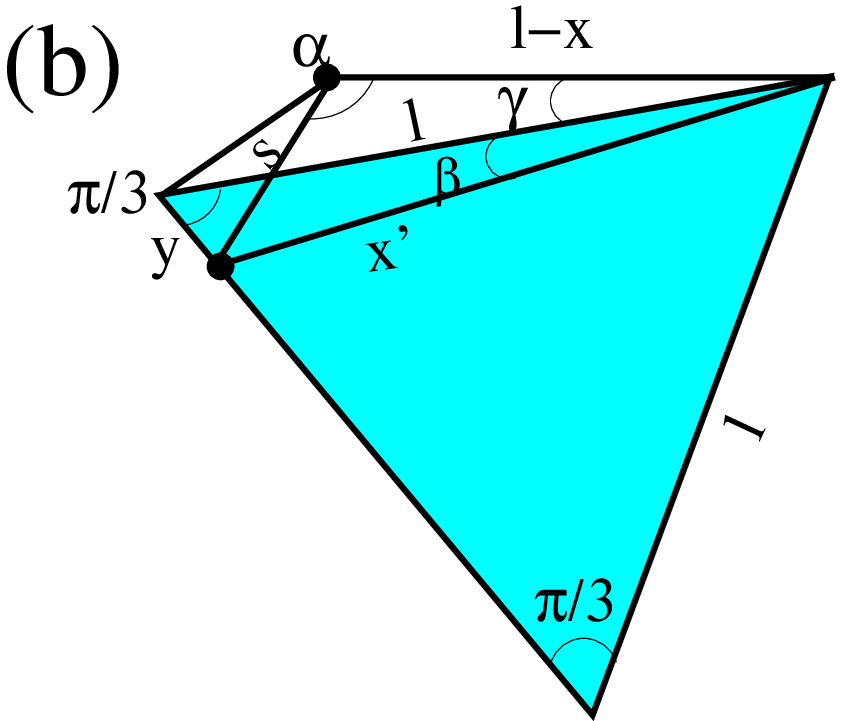,width=1.4in}\hspace*{0.5cm}
		\epsfig{file=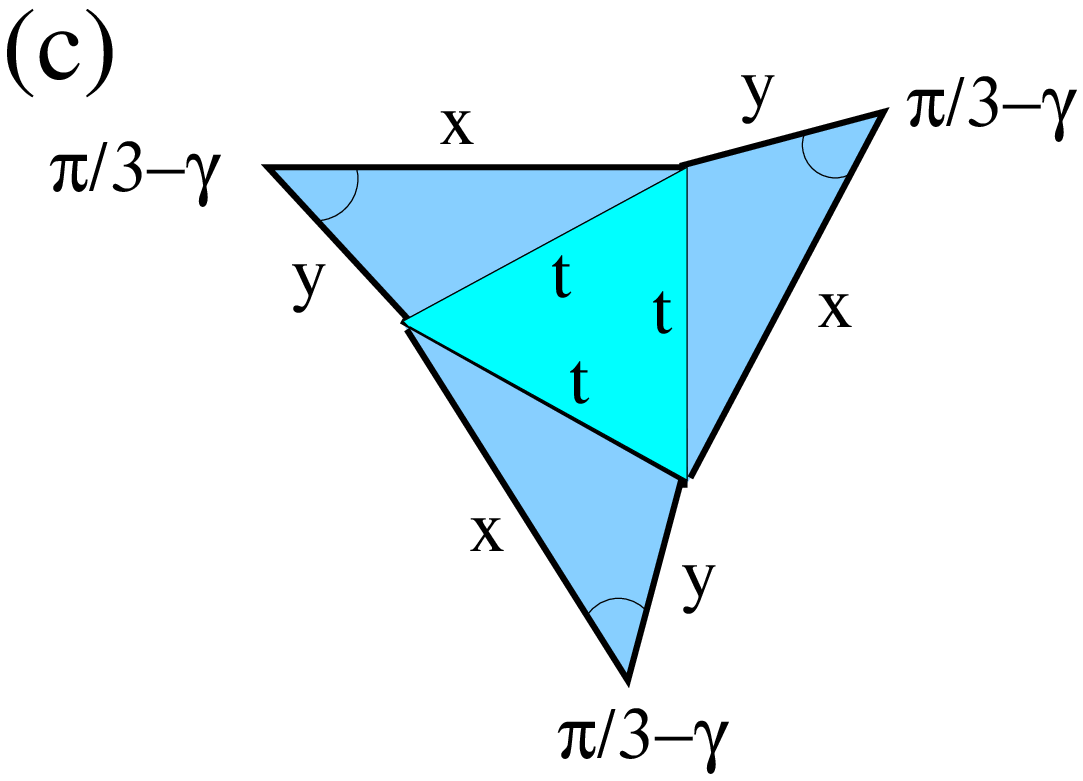,width=1.8in}
		\caption{(a) Sketch of the caged triangle rotated counterclockwise 
		in three different positions when in contact with three different pairs 
		of caging triangles for $\gamma<\gamma_-^{(1)}$. The free area is also shown.
		(b) Configuration where the caged triangle is in contact with two of its neighbours 
		(with their closest vertexes labeled with filled circles). The known quantities are the 
		angles $\alpha=2\pi/3+\theta$ and $\gamma$ and the lengths $l$ and $s$, while the unknowns 
		are $x'$, $\beta$, $x$ and $y$. 
		(c) Geometry of the free area for relatively small values of $\gamma$: 
		an equilateral triangle of size $t$ and three equal scalene 
		triangles of sizes $x$, $y$ and $t$, with angle $\pi/3-\gamma$ being the interior angle between
		the sides of lengths $x$ and $y$.}
		\label{fig4}
	\end{figure}

	In this section we show how to obtain the analytic expressions for the free area of a freely rotating 
	caged triangle inside the cell formed by its neighbours in the chiral triangular lattice configuration. 
	In Fig. \ref{fig4}(b) we show the caged triangle rotated counterclockwise in a configuration where it is in contact with two of its neighbours (two of their vertexes labeled with filled circles). The known quantities are the lengths $l$, $s$, and the angles $\gamma>0$ (the rotation angle) and $\alpha=2\pi/3+\theta$. The unknown quantities, necessary to find the free area, are $x$ and $y$. From the sinus-law of the triangle construction, shown in Fig. \ref{fig4}(b), we can find $x$ and $y$ as
	\begin{eqnarray}
		&&l-x=\frac{\sqrt{3}l/2-s\sin\left(\pi/3+\gamma+\theta\right)}{\sin\left(\pi/3-\gamma\right)}, \label{x}\\
		&&y=\frac{s\sin(\pi/3-\theta)-l\sin\gamma}{\sin\left(\pi/3-\gamma\right)}.\label{y}
	\end{eqnarray}
In Fig. \ref{fig4}(c) a sketch is shown of the free area for relatively small angles $\gamma$. It consists of 
an equilateral triangular core of length $t$ and three identical scalene triangles of sizes $x$, $y$ and $t$, 
with the interior angle between the sides of sizes $x$ and $y$ being equal to $\pi/3-\gamma$. The free area 
for this case can be computed as 
\begin{eqnarray}
	{\cal A}^{({\rm L},1)}(\gamma)=\frac{\sqrt{3}}{4}\left(x^2+y^2+4xy\cos(\pi/3+\gamma)\right),
\end{eqnarray}
which, after substituting Eqns. (\ref{x}) and (\ref{y}), becomes (in units of $a_0$)
\begin{eqnarray}
	&&\frac{{\cal A}^{({\rm L},1)}(\gamma)}{a_0}=\frac{1}{2}+2\tau\left[\cos(\psi)+\sqrt{3}
	\sin(\psi)\right]+\tau^2\left[1+\sqrt{3}\sin(2\psi)\right]\nonumber\\
	&&-4\left[1+\tau\cos(\psi)\right]\cos(\gamma')+4\tau \sin(\psi)\sin(\gamma')
	-\frac{6\tau\sin(\psi)}{\sin(\gamma')}\nonumber\\&& +
	\sqrt{3}\left[\frac{3}{2}+2\tau^2\sin^2(\psi)\right]\cot(\gamma'), \quad 0\leq \gamma\leq \gamma_-^{(1)}
	\label{al1}
\end{eqnarray}
where we have used the superindexes L and 1 to indicate that it is the first (1) part of the 
free area when the caged triangle is rotated to the left (L), i.e. counterclockwise. In (\ref{al1}) 
we have used the notations 
$\tau=s/l=\sqrt{\eta^{-1}-\sin^2\theta}-\cos\theta$, $\psi=\pi/3-\theta$, and $\gamma'=\pi/3-\gamma$, while 
$\gamma_-^{(1)}=\arcsin(\tau\sin\psi)$ is the angle for which the length $y$ becomes zero and the free area 
collapses to an equilateral triangle. Note that $\gamma_-^{(1)}=0$ for $\theta=\pi/3$. 
Integrating (\ref{al1}) with respect to $\gamma$ from 0 to $\gamma_-^{(1)}$ we obtain
\begin{eqnarray}
	&&\frac{A_{\rm free}^{({\rm L},1)}}{a_0}\equiv 
	\frac{\int_0^{\gamma_-^{(1)}} d\gamma {\cal A}^{({\rm L},1)}(\gamma)}{a_0}
	\nonumber\\
        &&=\left[\frac{1}{2}+2\tau\left(\cos\psi+\sqrt{3}\sin\psi\right)+\tau^2\left(1+\sqrt{3}\sin(2\psi)\right)
	\right]\gamma_-^{(1)}\nonumber\\
	&&+4(1+\tau\cos\psi)\left(\sin \delta_--\frac{\sqrt{3}}{2}\right)+4\tau\sin\psi\left(\cos \delta_--\frac{1}{2}\right)
	\nonumber\\
	&&+6\tau\sin\psi \ln\left[\sqrt{3}\tan(\delta_-/2)\right]+
	\sqrt{3}\left(\frac{3}{2}+2\tau^2\sin^2\psi\right)\ln\left[\frac{\sqrt{3}}{2}\csc \delta_-\right], \label{integrated1}
\end{eqnarray}
with $\delta_-=\pi/3-\gamma_-^{(1)}$.
\begin{figure}
	\epsfig{file=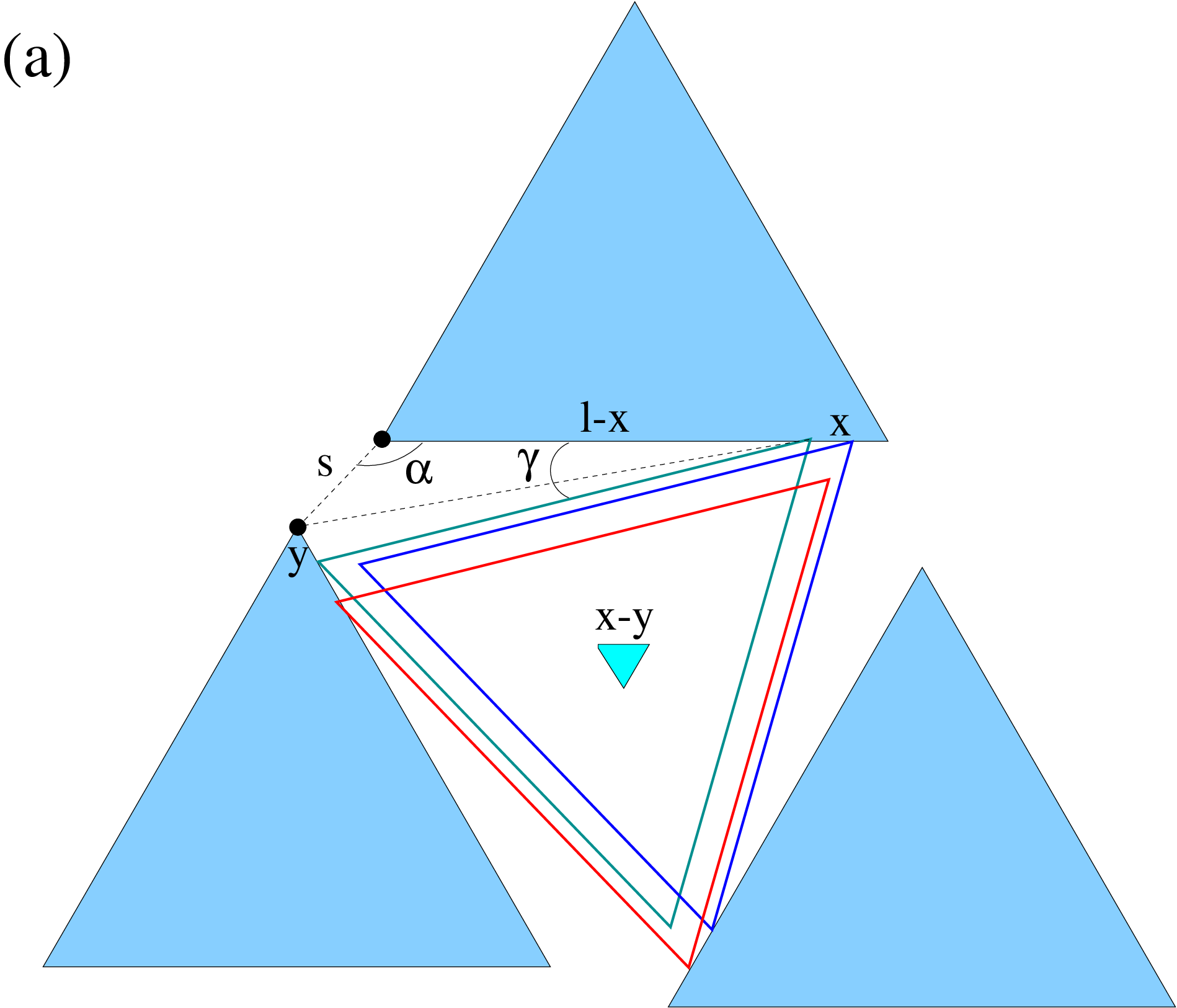,width=2.5in} \hspace*{0.5cm}
	\epsfig{file=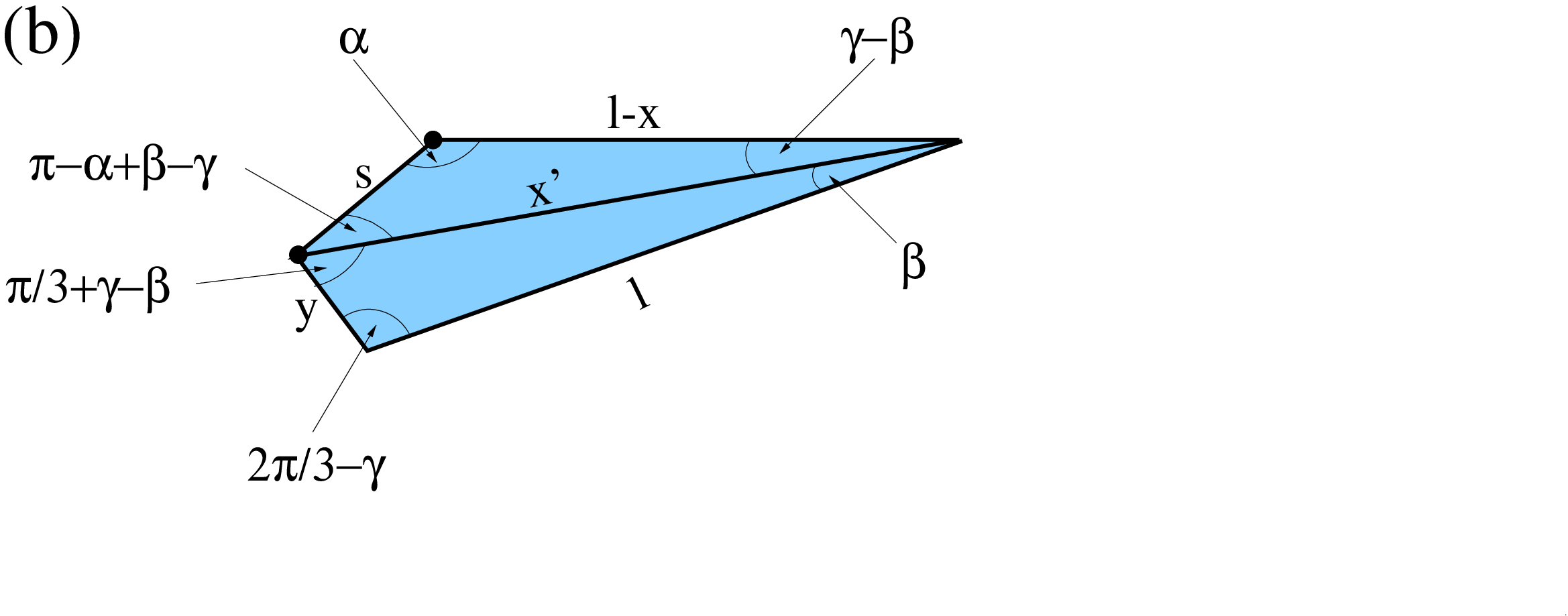,width=3.5in}
	\caption{(a) Sketch of the caged triangle rotated counterclockwise 
	in three different positions when in contact with three
		different pairs of caiging triangles for $\gamma>\gamma_-^{(1)}$. 
		(b) Geometric construction to find the variables $x$ and $y$, 
		necessary to calculate the free area. 
	Filled circles indicate the vertexes of the neighbours to the caged triangles. 
	The known quantities are $l$, $s$, $\alpha=2\pi/3+\theta$ and $\gamma$, while the unknowns 
	are $\beta$, $x'$, $x$ and $y$.}
	\label{fig5}
\end{figure}
For $\gamma>\gamma_-^{(1)}$ the main geometric construction to find the variables $x$ and $y$ is sketched in Fig. 
\ref{fig5} (b). This time the free area is an equilateral triangle of size $x-y$, so the free area 
is ${\cal A}^{({\rm L},2)}(\gamma)=\sqrt{3}\left(x-y\right)^2/4$. From Fig. \ref{fig5} we find:
\begin{eqnarray}
	&&l-x=\frac{2}{\sqrt{3}}\left[l\sin\left(\pi/3+\gamma\right)-s\sin\left(\pi/3+\theta\right)\right], \label{x2}\\
	&&y=\frac{2}{\sqrt{3}}\left[l\sin\gamma-s\sin(\pi/3-\theta)\right] \label{y2},
\end{eqnarray}
Thus we have
\begin{eqnarray}
	\frac{{\cal A}^{({\rm L},2)}(\gamma)}{a_0}=3+4\tau \cos\theta+4\tau^2\cos^2\theta
	-4(1+2\tau\cos\theta)\cos(\pi/3-\gamma)-2\cos(\pi/3+2\gamma),\nonumber\\
	\label{segunda}
\end{eqnarray}
valid for $\gamma_-^{(1)}\leq \gamma\leq \gamma_-^{(2)}$, where $\gamma_-^{(2)}$ is the angle for which $x-y=0$:
\begin{eqnarray}
	\gamma_-^{(2)}=\frac{\pi}{3}-\arccos\left(\frac{1}{2}+\tau\cos\theta\right).
\end{eqnarray}
The integration of (\ref{segunda}) with respect to $\gamma$ from $\tilde{\gamma}_-$ to $\gamma_-^{(2)}$, where 
$\tilde{\gamma}_-=\gamma_-^{(1)}$ for $\theta\leq \pi/3$ while $\tilde{\gamma}_-=0$ for $\theta>\pi/3$, gives
\begin{eqnarray}
	&&\frac{A_{\rm free}^{({\rm L},2)}}{a_0}=\frac{\int_{\tilde{\gamma}_-}^{\gamma_-^{(2)}}d\gamma {\cal A}^{({\rm L},2)}
	(\gamma)}{a_0}=\left(3+4\tau\cos\theta+4\tau^2\cos^2\theta\right)\left(\gamma_-^{(2)}-\tilde{\gamma}_-\right)
	\nonumber\\
	&&+4(1+2\tau\cos\theta)\left(\sin\left(\frac{\pi}{3}-\gamma_-^{(2)}\right)-\sin\left(\frac{\pi}{3}
	-\tilde{\gamma}_-\right)\right)+\sin\left(\frac{\pi}{3}+2\tilde{\gamma}_-\right)
	-\sin\left(\frac{\pi}{3}+2\gamma_-^{(2)}\right).\label{integrated2} \nonumber\\
\end{eqnarray}
Thus the total integrated free area for counterclockwise rotation can be found from
\begin{eqnarray}
	A_{\rm free}^{({\rm L})}=\left\{
		\begin{matrix}
			3\left(A_{{\rm free}}^{({\rm L},1)}+A_{\rm free}^{({\rm L},2)}\right), & \text{if}\ \displaystyle{
				0\leq\theta\leq \frac{\pi}{3}},\\
			3A_{\rm free}^{({\rm L},2)}, & \text{if}\ \displaystyle{\frac{\pi}{3}\leq \theta\leq \frac{\pi}{2}.}
		\end{matrix}
		\right.
		\label{ultima}
\end{eqnarray}
where the factor 3 is due to the presence of three sectors of equal areas due to the symmetry of the cell geometry.

\begin{figure}
	\epsfig{file=fig11a.eps,width=3.in}
	\epsfig{file=fig11b.eps,width=3.in}
	\caption{The left free area $A_{\rm free}^{({\rm L})}/a_0$ as a function of (a) $\theta$, and (b) $\phi$,
	measured in degrees for three different values (labeled) of $\eta$.}
	\label{fig6}
\end{figure}

Fig. \ref{fig6} plots the function $A_{\rm free}^{({\rm L})}(\theta)$ as a function of $\theta$ (a) or $\phi$ (b)
for three different values of $\eta$. 
Clearly it is a decreasing function of $\theta$.

When we rotate the caged triangle clockwise (in the same direction of chirality), the free areas are different. For small enough angle $\gamma$, the triangular constructions to find the magnitudes $x$ and $y$ are similar to that given in Fig. \ref{fig4}(b) and Fig. \ref{fig5} (b), but with an important difference. Now we have 
$\alpha=2\pi/3-\theta$ (instead of $2\pi/3+\theta$), so the magnitudes $x$ and $y$ are given by the same Eqns. (\ref{x})-(\ref{y}) and (\ref{x2})-(\ref{y2}) except for the substitution $\theta \to-\theta$. Thus we have that the free areas have the same expressions as in Eqns. (\ref{al1}) and (\ref{segunda}) with this substitution, i.e. 
\begin{eqnarray}
	&&{\cal A}^{({\rm R},1)}(\gamma;\theta)={\cal A}^{({\rm L},1)}(\gamma;-\theta), \ \text{for} \ 0\leq \gamma\leq \gamma_+^{(1)} \label{solo}\\ 
	&&{\cal A}^{({\rm R},2)}(\gamma;\theta)={\cal A}^{({\rm L},2)}(\gamma;-\theta), \ \text{for} \ \gamma_+^{(1)}\leq \gamma \leq \gamma_+^{(2)},
\end{eqnarray}
where we have 
\begin{eqnarray}
	\gamma_+^{(i)}(\theta)=\gamma_-^{(i)}(-\theta), \quad i=1,2.
\end{eqnarray}
\begin{figure}
	\epsfig{file=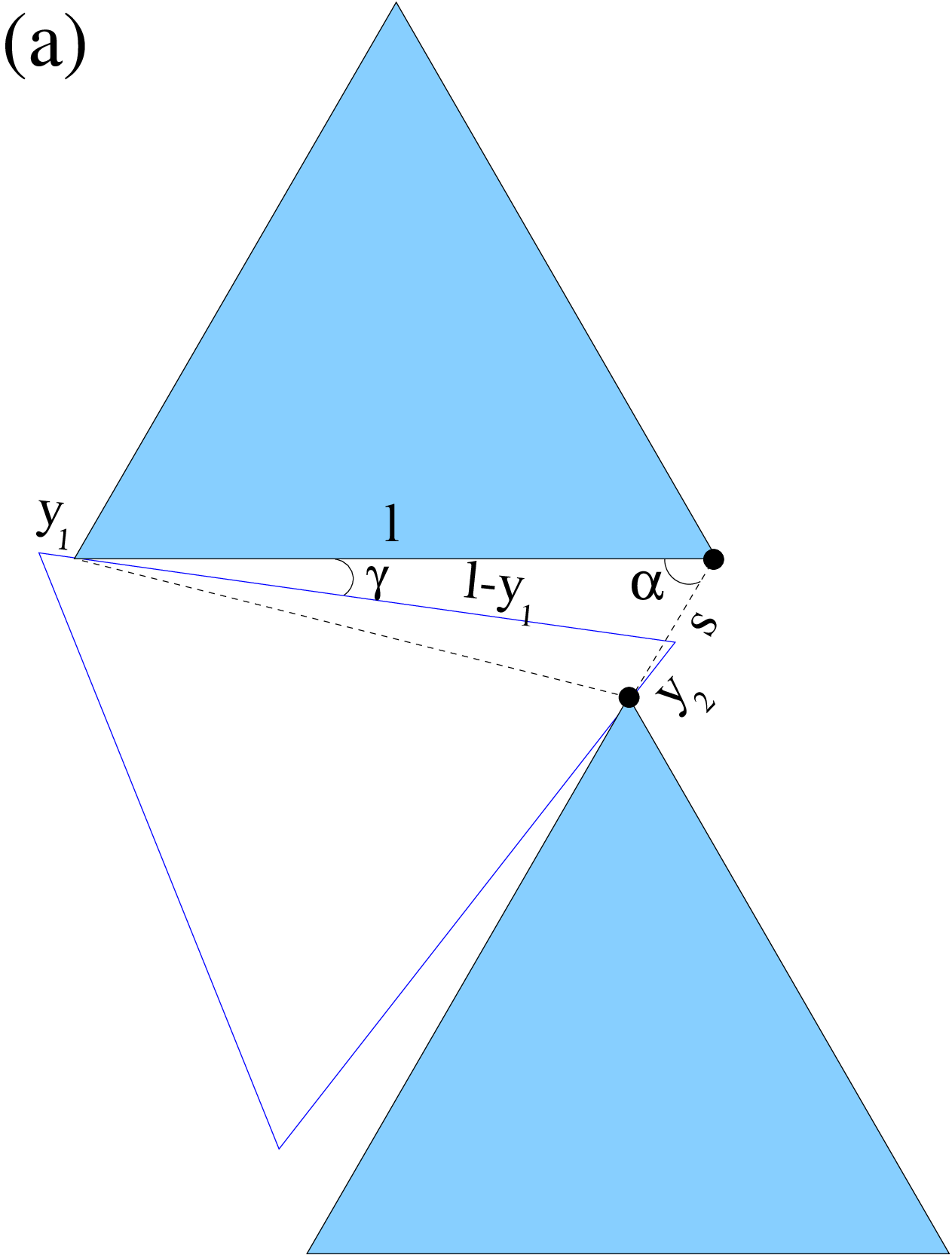,width=2.in}\hspace*{1.cm}
	\epsfig{file=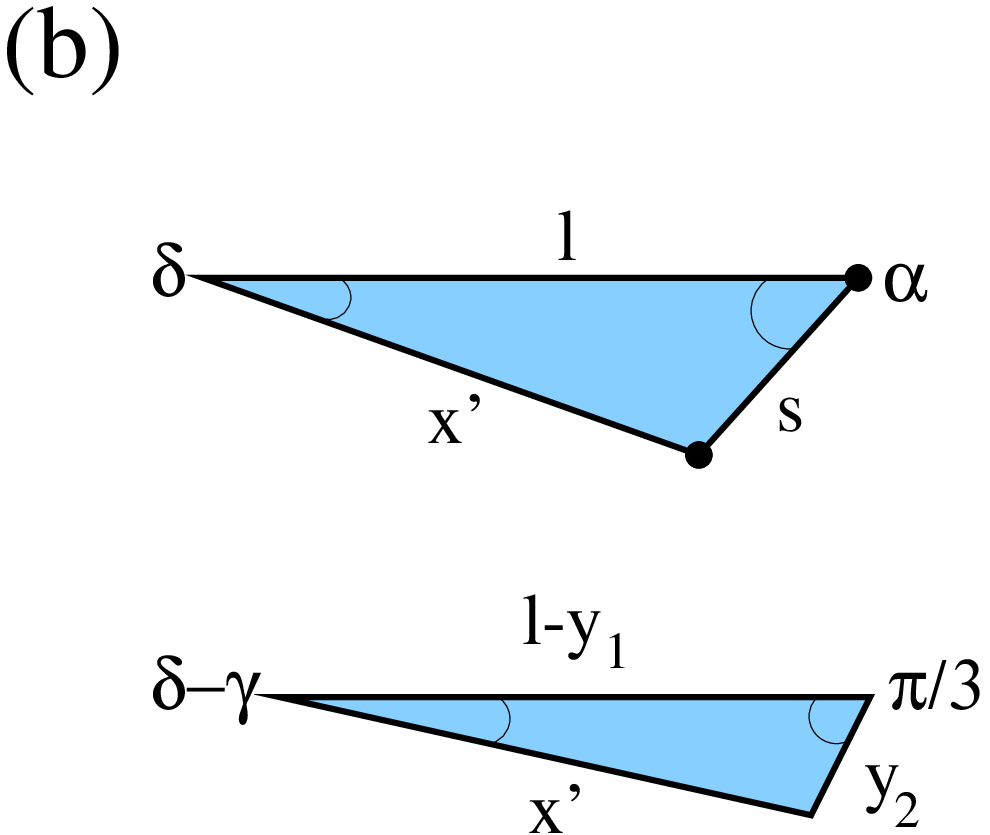,width=2.in}
	\caption{(a) Sketch of the caged triangle rotated clockwise in contact with two caging triangles for $\theta>\theta^*$ 
	and $\gamma>\gamma_{+}^{(3)}$. 
	The free area (not shown) is an equilateral triangle of size $y_2-y_1$.
	(b) Triangle constructions from which the magnitudes $y_1$ and $y_2$, and consequently the free area, can be found. 
	The lengths $l$ and $s$ and the angles $\alpha$ and $\gamma$ are known, while the lengths $x'$, $y_1$ and $y_2$ 
	and the angle $\delta$ are unknown.}
	\label{fig7}
\end{figure}
But now the length $x$ from Eqn. (\ref{x}) with $\theta\to-\theta$ can be equal to zero (which in turn implies that the free area becomes a equilateral triangle of side $y$), and this takes place at an angle 
\begin{eqnarray}
	\gamma_+^{(3)}=\arcsin\left[\frac{\sqrt{3}/2+\tau\sin(\pi/3-\theta)}{{\cal L}}\right]
	-\arcsin\left[\frac{\sqrt{3}/2}{{\cal L}}\right].
\end{eqnarray}
where we have defined the quantity
\begin{eqnarray}
	{\cal L}=\sqrt{1+2\tau\cos(\pi/3+\theta)+\tau^2}.
\end{eqnarray}
The equality $\gamma_+^{(1)}=\gamma_+^{(2)}=\gamma_+^{(3)}$ takes place at 
$\displaystyle{\theta=\theta^*=\arcsin\left(\frac{1}{2\sqrt{\eta}}\right)}$. 
Thus, for $\theta\leq\theta^*$, the integrated free areas $A_{\rm free}^{({\rm R},i)}$ ($i=1,2$) has the same expressions as those from Eqns. (\ref{integrated1}) and (\ref{integrated2}), except for the substitutions $\theta\to-\theta$ and $\gamma_-^{(i)}\to\gamma_+^{(i)}$. 
For $\theta>\theta^*$ and $0\leq \gamma\leq \gamma_+^{(3)}$, we have again the equality (\ref{solo}) with the substitution 
$\gamma_+^{(1)}\to \gamma_+^{(3)}$, necessary to calculate the integrated area $A_{\rm free}^{({\rm R},1)}$ in this interval of angles. However, for $\gamma>\gamma_+^{(3)}$,   
we obtain that the free area is an equilateral triangle of length $y_1-y_2$, where the magnitudes $y_i$ can be 
calculated from the triangular construction shown in Fig. \ref{fig7} (b), with the result
\begin{eqnarray}
	&&y_1=l\left[1-\frac{2}{\sqrt{3}}\sin(\pi/3-\gamma)\right]-\frac{2}{\sqrt{3}}s\sin(\pi/3+\gamma-\theta),\\
	&&y_2=\frac{2}{\sqrt{3}}\left[s\sin(\pi/3+\theta-\gamma)-l\sin\gamma\right].
\end{eqnarray}
Thus we obtain the free area
\begin{eqnarray}
	&&\frac{{\cal A}^{({\rm R},2)}(\gamma)}{a_0}=3+4\tau\cos(\pi/3+\theta)+2\tau^2\nonumber\\
	&&+\left[4\tau\cos(\pi/3-\theta)+2\tau^2\cos(2\theta)-1\right]\cos(2\gamma)\nonumber\\
	&&-2(1+2\tau\cos\theta)\cos\gamma+2\left(\sqrt{3}-2\tau\sin\theta\right)\sin\gamma\nonumber\\
	&&-2\left[\frac{\sqrt{3}}{2}+2\tau\sin(\pi/3-\theta)-\tau^2\sin(2\theta)\right]\sin(2\gamma),
	\label{otro}
\end{eqnarray}
which has the primitive 
\begin{eqnarray}
	&&{\cal S}(\gamma)=\int d\gamma \frac{{\cal A}^{({\rm R},2)}(\gamma)}{a_0}=3\gamma+\sin(\pi/3-2\gamma)-4\sin(\pi/3+\gamma)\nonumber\\
	&&+2\tau\left[2\gamma \cos(\pi/3+\theta)+\sin(\pi/3-\theta+2\gamma)+2\sin(\theta-\gamma)\right]\nonumber\\
	&&+\tau^2\left[2\gamma-\sin(2(\theta-\gamma))\right].
\end{eqnarray}
The excluded area (\ref{otro}) becomes zero when $y_1=y_2$, a condition that takes place at the angle 
\begin{eqnarray}
	\gamma_+^{(4)}=\arccos\left(\frac{1/2}{{\cal L}}\right)
	-\arccos\left(\frac{1/2+\tau\cos\theta}{{\cal L}}\right).
\end{eqnarray}
Note that $\gamma_+^{(3)}=0$ for $\theta=\pi/3$ while $\gamma_+^{(4)}=0$ for $\theta=\pi/2$. 
Then we have that the second integrated area for $\theta>\theta^*$ can be calculated from 
\begin{eqnarray}
	\frac{A_{\rm free}^{({\rm R},2)}}{a_0}=\left\{
		\begin{matrix}
			{\cal S}\left(\gamma_+^{(4)}\right)-{\cal S}\left(\gamma_+^{(3)}\right),& \displaystyle{
				\theta^*\leq\theta\leq \frac{\pi}{3}},\\
			{\cal S}\left(\gamma_+^{(4)}\right)-{\cal S}(0), & \displaystyle{\frac{\pi}{3}\leq \theta\leq \frac{\pi}{2}}.
		\end{matrix}
		\right.
\end{eqnarray}

\begin{figure}
	\epsfig{file=fig13a.eps,width=3.in}
	\epsfig{file=fig13b.eps,width=3.in}
	\caption{The right free area as a function of (a) $\theta$, and (b) $\phi$ (both in degrees) 
	for three different (labeled) values of packing fraction. }
	\label{fig8}
\end{figure}

Finally, the total area for clockwise rotation is given again by the same Eqn. (\ref{ultima}), and it 
is plotted in Fig. \ref{fig8} as a function of $\theta$ (a) or $\phi$ (b) for three different values of $\eta$. Note 
that they exhibit small maxima at angles close, but not equal, to $0$. However the total free area, $A_{\rm free}=A_{\rm free}^{({\rm L})}+A_{\rm free}^{({\rm R})}$, exhibits a monotonic behavior, as seen in  
Fig. \ref{fig1}. Therefore, the presence of these maxima are not enough to generate a non-monotonic behavior of the total free-area (which would explain the presence of chirality in MC simulations).

\begin{figure}
	\epsfig{file=fig14a.eps,width=3.in}
	\epsfig{file=fig14b.eps,width=2.9in}
	\caption{(a) The functions ${\cal A}^{({\rm L})}(\gamma)$ (solid) and 
	${\cal A}^{({\rm R})}(\gamma)$ (dashed) for different values of $\theta$ as they labeled and 
	for a fixed value of $\eta=0.8$. With different symbols we show that maximum values of $\gamma$ for each curve. (b) The equilibrium angular distribution functions $f_{\theta}(\gamma)$ for the same values of $\theta$ and $\eta$. }
	\label{fig9}
\end{figure}

The reason for the nonmonotonic behavior of $A_{\rm free}^{({\rm R})}$ as a function of $\theta$ or $\phi$ 
can be easily explained by examining the functions ${\cal A}^{(\nu)}(\gamma)={\cal A}^{(\nu,1)}(\gamma)+{\cal A}^{(\nu,2)}(\gamma)$ ($\nu=\left\{L,R\right\}$) for different values of $\theta$, see Fig. \ref{fig9} (a). It can be seen that, for certain values of $\theta_2>\theta_1$, the inequality 
${\cal A}^{({\rm R})}_{\theta_2}(\gamma)>\left\{{\cal A}^{({\rm L})}_{\theta_1}(\gamma),
{\cal A}^{({\rm R})}_{\theta_1}(\gamma)\right\}$ is fulfilled for high enough values of $\gamma$. Thus, there exist values of $\theta_1,\theta_2\sim 0$ for which the inequality between the integrals 
$A_{{\rm free},\theta_2}^{({\rm R})}=3\int d\gamma {\cal A}^{({\rm R})}_{\theta_2}(\gamma)>A_{{\rm free},\theta_1}^{({\rm R})}=3\int d\gamma {\cal A}^{({\rm R})}_{\theta_1}(\gamma)$  
is also fulfilled which, in turn, explains the presence of maxima 
of $A_{\rm free}^{({\rm R})}$ as a function of $\theta$ 
or $\phi$ at a vanishing angle. Fig. \ref{fig9}(b) is a plot of the equilibrium angular 
distribution function $f_{\theta}(\gamma)$  
calculated from Eqn. (\ref{equil}) with ${\cal A}(\gamma)={\cal A}^{({\rm L})}(|\gamma|)$ if $\gamma<0$, and 
${\cal A}(\gamma)={\cal A}^{({\rm R})}(|\gamma|)$ if $\gamma\geq 0$. It can be seen that it is not symmetric with respect to $\gamma=0$. This in turn implies that the 
value of $\langle \gamma \rangle\equiv \int_{-\gamma_-}^{\gamma_+} d\gamma  \gamma f_{\theta}(\gamma) \neq 0$.
We obtain $\langle \gamma\rangle=0.7130^{\circ}$, $0.3815^{\circ}$, $0.1561^{\circ}$ and $0.0405^{\circ}$ for 
$\theta=75^{\circ}$, $60^{\circ}$, $45^{\circ}$ and $30^{\circ}$, respectively.

\acknowledgements

Financial support from Grant No. PID2021-126307NB-C21 (MCIU/AEI/FEDER,UE) is acknowledged.

\end{appendix}

\end{document}